\documentclass[a4paper,11pt]{article}
\usepackage{jcappub} 

\usepackage{enumitem}
\usepackage{fancyhdr}
\usepackage{mathtools}
\usepackage{subfig}
\usepackage{amssymb}
\usepackage{amsmath}
\usepackage{hyperref}
\usepackage{amsfonts}
\usepackage{amssymb}
\usepackage{caption}
\usepackage{float}
\usepackage{bm}
\usepackage{graphicx}
\usepackage{color}
\usepackage{verbatim}
\usepackage{blindtext, multicol}
\usepackage{multirow}
\usepackage{epigraph}
\usepackage[dvipsnames]{xcolor}
\usepackage{physics}

\usepackage[normalem]{ulem}
\newcommand\Tstrut{\rule{0pt}{2.9ex}} 

\usepackage[capitalise]{cleveref}
\usepackage[activate={true,nocompatibility},final,kerning=true,factor=1100,stretch=10,shrink=10]{microtype}

\usepackage[sorting=none,style=numeric-comp]{biblatex}
\usepackage[english]{babel}
\usepackage{csquotes}
\DeclareUnicodeCharacter{0301}{\'{i}}
\usepackage[nottoc]{tocbibind}

\usepackage{academicons}
\usepackage{xcolor}
\usepackage{fontawesome5}
\definecolor{orcidlogocol}{rgb}{0.65, 0.807, 0.223}

\newcommand{\orcid}[1]{$\,$\href{https://orcid.org/#1}{\textcolor{orcidlogocol}{\faOrcid}}}

\usepackage[T1]{fontenc} 

\def\beq{\begin{equation}}
\def\eeq{\end{equation}}
\def\ber{\begin{eqnarray}}
\def\eer{\end{eqnarray}}
\def\benu{\begin{enumerate}}
\def\eenu{\end{enumerate}}

\def\Tre{T_{_{\rm re}}}

\def\wre{w_{_{\rm re}}}
\def\Nre{N_{_{\rm re}}}
\def\l{\left}
\def\r{\right}

\newcommand{\sq}{\lower.25ex\hbox{\large$\Box$}}
\def\n{\nabla}

\def\f{\frac}
\def\mpl{m_{p}}

\def\mpc{\rm Mpc^{-1}}
\def\ns{n_{_S}}
\def\nt{n_{_T}}

\def\ng{n_{_{\rm GW}}}
\def\Og{\Omega_{_{\rm GW}}}

\def \lleq {\lower0.9ex\hbox{ $\buildrel < \over \sim$} ~}
\def \ggeq {\lower0.9ex\hbox{ $\buildrel > \over \sim$} ~}

\def\n {\noindent}

\hypersetup{
    colorlinks = true,
    linkcolor = blue,
    citecolor = blue,
    urlcolor = Blue
}

\title{ New models of Quintessential Inflation featuring  plateau and hilltop potentials}
\author[a]{Swagat S. Mishra \orcid{0000-0003-4057-145X},}
\author[b]{ Varun Sahni,}

\affiliation[a]{School of Physics and Astronomy,  University of Nottingham, Nottingham, NG7 2RD, UK.}
\affiliation[b]{Inter-University Centre for Astronomy and Astrophysics,
Post Bag 4, Ganeshkhind, Pune 411~007, India}

\emailAdd{swagat.mishra@nottingham.ac.uk}
\emailAdd{varun@iucaa.in}

\abstract{We introduce a new class of hilltop and plateau potentials which can
successfully unify inflation and dark energy resulting in Quintessential Inflation (QI).
Interestingly these new potentials are related through an inverse transformation.
Namely, if $V(\phi) = V_0 \, v(\phi)$ is a plateau potential then the inverse potential
$V(\phi) = V_0 \left[v(\phi)\right]^{-1}$ describes hilltop QI. 
A simple example is provided by the KKLT-inspired potential
$v(\phi) = \left\lbrack \frac{M^{2n} + \phi^{2n}}{N^{2n} + \phi^{2n}}\right\rbrack \,$.
When $M/N \ll 1$ this potential describes plateau QI, while its inverse, $\left[v(\phi)\right]^{-1}$ 
describes hilltop QI.
Other simple models of QI arise for the class of potentials
$V(\phi) \sim \exp\left\lbrack\mp f(\phi)\right\rbrack$,
where the $-$ ($+$) sign is associated with a plateau (hilltop).
A key feature of this new class of QI models 
is the near absence of small parameters which are usually associated with the presence of 
dark energy. 
A forecast for the gravitational wave background generated in these
models is provided.
} 

\begin{document}
\maketitle

\section{Introduction}
\label{sec:intro}
A remarkable aspect our universe is that it appears to accelerate twice, once at very early times 
during `inflation' \cite{Starobinsky:1979ty,Starobinsky:1980te,Guth:1980zm,Linde:1981mu,Albrecht:1982wi,Linde:1983gd,Baumann_TASI,Baumann:2018muz,Baumann:2022mni,Martin:2013tda}, and again at very late times, close to the present epoch.
In most theoretical studies the mechanism governing early acceleration (inflation) is
held to be quite distinct from that which causes late-time acceleration. 
This late-acceleration epoch is assumed to arise because of the presence of a new
degree of freedom called `dark energy' (DE). Like inflation, DE is assumed to possess a strongly
negative equation of state (EOS), $w = p/\rho < -1/3$, with the value $w=-1$ being associated
with the famous cosmological constant \cite{Sahni:1999gb,Peebles:2002gy,Copeland:2006wr,Sahni:2006pa}. 
The reason why Inflation and DE are considered to be
distinct is not difficult to see. 
Inflation takes place at very high energies not much lower than the Planck scale, 
whereas DE begins to dominate the universe much later when its expansion rate $H = \dot{a}/a$ has slowed
considerably, and is much smaller than the inflationary value.
Therefore it is usually assumed that the fundamental parameters which enter into inflationary
model building must be several orders of magnitude larger than those defining dark energy.

Despite these issues, the idea of unifying the early and the late-time acceleration
has considerable appeal and numerous efforts have been made to construct
viable models of Quintessential Inflation (QI); see \cite{Peebles:1998qn,Peloso:1999dm,Dimopoulos:2000md,Majumdar:2001mm,Sami:2004xk,Dimopoulos:2001ix,Giovannini:2003jw,Rosenfeld:2005mt,Hossain:2014xha,Hossain:2014coa,WaliHossain:2014usl,Dimopoulos:2017zvq,Rubio:2017gty,Ahmad:2017itq,Geng:2017mic,vandeBruck:2017voa,Haro:2019gsv,Rosa:2019jci,Haro:2019peq,Kleidis:2019ywv,Lima:2019yyv,Benisty:2020nuu,Benisty:2020xqm,Copeland:2000hn,Dimopoulos:2021xld,Jaman:2022bho} and references therein.

QI usually relies on a single field (usually a scalar) to unify inflation and dark energy.
Since the field connects physics at vastly different energy scales, it follows that
the potential $V(\phi)$ giving rise to QI should be large at early times 
and small at the present epoch. Moreover, if $\phi$ is the sole field giving rise to both inflation and DE
then its potential $V(\phi)$ should be sufficiently flat both at early and late times.
This condition is required so that $\phi$ can violate the strong energy condition (SEC)
$\rho_\phi + 3p_\phi \geq 0$ and give rise to cosmic acceleration.\footnote{In modified
gravity theories such as braneworld models \cite{Randall:1999vf,Randall:1999ee}, the requirement that the potential is initially
flat is no longer necessary since corrections to the Einstein-Hilbert action can ensure
slow-roll even for a steep potential \cite{Maartens:1999hf,Copeland:2000hn}.}
A further requirement on $V(\phi)$ is that it allows for post-inflationary reheating to occur.
Conventional mechanisms for reheating are usually based on either one of the following scenarios:

\begin{enumerate}

\item The rapidly changing gravitational field, post inflation, gives rise to quantum
particle production which reheats the universe (gravitational reheating) \cite{Ford:1986sy,Spokoiny:1993kt,Kolb:2023ydq}.

\item Scalar field oscillations around a minimum of the inflationary potential source reheating either
through the perturbative or non-perturbative decay of the inflaton \cite{Bassett:2005xm,Allahverdi:2010xz,Kofman:1996mv,Kofman:1994rk,Kofman:1997yn,Shtanov:1994ce}. 

\item In the absence of a minimum in the potential, reheating can still occur if the potential
declines very rapidly after inflation. The steep decline in $V(\phi)$ leads to a rapid
growth in ${\dot\phi}^2$ which can result in `instant preheating' \cite{Felder:1998vq,Felder:1999pv}.


\end{enumerate}

The gravitational reheating (mechanism 1) is quite inefficient and  usually results in a  long duration of reheating with a low reheating temperature.
For potentials exhibiting a stiff EOS post inflation \cite{Copeland:2000hn,Apers:2024ffe},  the  long duration of
the kinetic regime leads
to an overproduction of
 post-inflationary gravitational waves 
 whose density can violate the Big Bang  Nucleosynthesis bound
$\left (\rho_g/\rho_r\right ) \lleq 0.2$ (as shown in \cite{Sahni:2001qp}).
Consequently this method of reheating is now seldom used in constructing models of QI
and most current models of Quintessential Inflation usually rely on 
mechanisms 2 or 3 to reheat the universe.

\medskip

In this paper we propose several new models of QI.
\begin{itemize}

\item
Our first model is based on a plateau potential with two symmetric flat wings.
In this model inflation occurs during the slow-roll of the inflaton along one of the wings.
A key feature of this model is that the potential does not vanish at its minimum
value. Instead $V(\phi=0) > 0$ implying that the ground state
in this model has a positive energy which leads to $\Lambda$CDM-like expansion
at late times. Moreover, since $V''(\phi=0) > 0$ scalar field oscillations
around $\phi = 0$ can reheat the universe 
as in conventional inflationary models.

\item
The second model of QI is sourced by a hilltop potential with flat wings.
The hilltop model has the following properties:

\begin{enumerate}[label=(\roman*)]

\item $V(\phi)$ is flat at the origin
which facilitates slow-roll
inflation.

\item $V(\phi)$ is also flat at its wings with
$V(\phi\to \pm\infty) = V_0$,
 allowing for cosmic acceleration to recur
close to the present epoch.

\item In between these two flat regions the potential steepens considerably. This is required
(a) to ensure that $V(\phi)$ drops from its initially high inflationary value to the tiny
value associated with dark energy, and
(b) to ensure that the post-inflationary universe can reheat via the instant preheating
mechanism \cite{Felder:1998vq,Felder:1999pv} which works well for steep potentials.

\end{enumerate}

\end{itemize}
A key feature of the plateau and hilltop models which we propose is the absence of small parameters which are usually present in models of
dark energy. Indeed, although 
 DE takes the form of a cosmological
constant $V_0\equiv \Lambda$, the tiny value of $\Lambda$ in our models
is associated with a much larger energy scale already present
in the inflationary potential.
Moreover the plateau and hilltop models of QI are related through an inverse transformation.
Namely, if $V(\phi) = V_0 \, v(\phi)$ corresponds to a plateau potential then the inverse potential
$V(\phi) = V_0 \left[v(\phi)\right]^{-1}$ describes a hilltop with flat wings.
Both potentials provide compelling models of quintessential inflation.

\bigskip

Our paper is organised as follows: In section \ref{sec:QI_dual_KKLT} we demonstrate that
two simple models of Quintessential Inflation can arise from one and the same potential. 
The first is a plateau model, discussed in section \ref{sec:QI_plateau_KKLT}, while the second
is a flat-winged hilltop, discussed in section \ref{sec:QI_hilltop_KKLT}.  We discuss
two more  models of QI featuring exponential-based potentials in section  \ref{sec:QI_dual_exp}.  
In section \ref{sec:QI_nonflat} we introduce several other models of QI having different properties. Section \ref{sec:QI_inf_dual} is dedicated to the analysis of  inflationary dynamics of  plateau and hilltop QI potentials. 
After a brief discussion of reheating in section \ref{sec:QI_reheating_dual}, we study the associated spectrum of relic  gravitational waves  in section \ref{sec:GW}. 
Section \ref{sec:discussion} contains a discussion of our main results. 

\medskip

We assume the background
 Universe to be described by a spatially flat Friedmann-Lemaitre-Robertson-Walker (FLRW)  metric with signature $(-,+,+,+)$. We work in natural units with $c=\hbar =1$ and define the reduced Planck mass to be  $\mpl \equiv 1/\sqrt{8\pi G} = 2.43 \times 10^{18}~{\rm GeV}$.

\section{Plateau and hilltop potentials for quintessential inflation}
\label{sec:QI_dual_KKLT}

Two simple models of quintessential inflation can be generated by the pair of
potentials
\beq
V(\phi) = V_0\,v(\phi) ~~~{\rm and} ~~ V(\phi) = V_0\,\left[v(\phi)\right]^{-1}~,
\label{eq:plateau_hilltop}
\eeq
where
\beq
v(\phi) = \left\lbrack \frac{M^{2n} + \phi^{2n}}{N^{2n} + \phi^{2n}}\right\rbrack~. 
\label{eq:plateau_hilltop_KKLT}
\eeq
For $M \ll N$ the potential $V(\phi) = V_0\,v(\phi)$ describes a plateau model of QI whereas its inverse
$V(\phi) = V_0\,\left[v(\phi)\right]^{-1}$ describes hilltop QI.
This can be seen from the following considerations.

\subsection{Plateau QI}
\label{sec:QI_plateau_KKLT}

\begin{figure}[htb]
\centering
\includegraphics[width=0.9\textwidth]{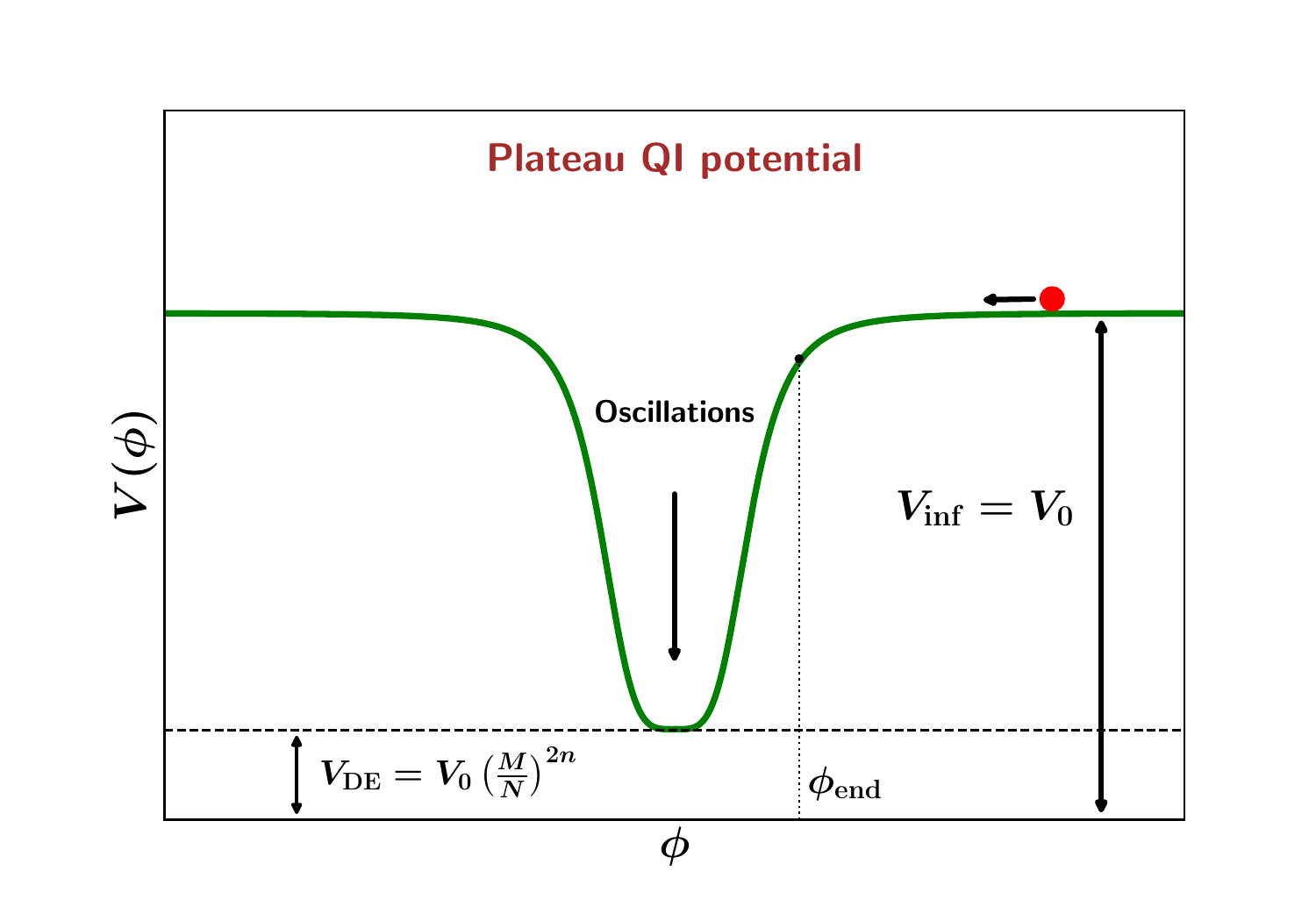}
\caption{The plateau potential given in Eq.~(\ref{eq:plateau_KKLT}) for quintessential inflation  is schematically depicted in this figure.}
\label{fig:pot_plateau_KKLT}
\end{figure}

The plateau potential sourcing QI
\beq
V_{\rm pl}(\phi) = V_0 \left\lbrack \frac{M^{2n} + \phi^{2n}}{N^{2n} + \phi^{2n}}\right\rbrack~,  ~~M \ll N
\label{eq:plateau_KKLT}
\eeq
has two important asymptotic limits (see Fig.~\ref{fig:pot_plateau_KKLT}):

\begin{enumerate}

\item For large values $\phi \gg \lbrace M,N\rbrace$ the potential approaches a constant
value $V(\phi) \to V_0$.

\item In the opposite case, at the origin, the potential approaches
\beq
V_{\rm plateau}(\phi = 0) = V_0\,v(0) = V_0\left (\frac{M}{N}\right )^{2n}~.
\label{eq:origin0}
\eeq

\end{enumerate}

One therefore finds that 
the flat wings of the plateau potential with $V(\phi\to\pm\infty)=V_0$ can be easily associated with inflation,
whereas the ground state
\beq
V_{\rm pl}(\phi = 0) = V_0\left (\frac{M}{N}\right )^{2n} \ll V_0~, ~~M \ll N
\label{eq:origin1}
\eeq
can give rise to dark energy. Indeed, substituting the CMB constraint $V_0 \simeq 10^{-10}m_p^4$ and
\beq
\rho_{_{\rm DE}} \equiv V_{\rm pl}(\phi=0) \simeq 10^{-120}m_p^4 
\eeq
into Eq.~(\ref{eq:origin0}) one finds
\beq
\left (\frac{M}{N}\right )^{2n} \sim 10^{-110} 
\label{eq:ratio_mM}
\eeq
The values $1\leq n \leq 4$, $N \sim m_p$ satisfy
CMB bounds, as discussed in section \ref{sec:QI_inf_dual_KKLT}.
One therefore  finds $M \sim GeV$ for $n=3$. Thus the exceedingly small DE value
$\rho_{\rm DE} \sim \left (10^{-3}\, eV\right )^4$ can be associated with fairly large values
of the constants $M$ and $N$ in this model of QI.

Note that in the plateau model inflation ends when $\phi \sim m_p$.
After that the scalar field begins to oscillate rapidly about the minimum of its potential
enabling the universe to reheat. Prior to reheating however, the equation of state
(EOS) of the oscillating scalar field is
\beq
\langle w_\phi\rangle = \frac{n-1}{n+1}
\label{eq:EOS}
\eeq
and the expansion of the universe is approximately described by
\beq
a(t) \propto t^\frac{2}{3(1+\langle w_\phi\rangle)}~.
\eeq
Since $1 \leq n \leq 4$ agree with CMB measurements (see Fig.~\ref{fig:KKLT_r_ns}), it follows that
the EOS of the scalar field prior to reheating lies in the interval $0 \leq w_\phi \leq 0.6$.
This has important implications for
the relic gravitational wave background, as discussed in section \ref{sec:Omega_GW}.

Finally note that the potential in Eq.~(\ref{eq:plateau_KKLT})
 has two important limits: for $M=0$ it reduces to the well known
KKLT potential associated with inflation \cite{Kachru:2003aw,Kachru:2003sx,Kallosh:2019jnl},
whereas for $N=0$, Eq.~(\ref{eq:plateau_hilltop_KKLT}) describes a tracker model of DE
which approaches a
cosmological constant at late times. Tracker models of DE with similar properties have been
 discussed in \cite{Bag:2017vjp}.

\subsection{Hilltop QI}
\label{sec:QI_hilltop_KKLT}

\begin{figure}[htb]
\centering
\includegraphics[width=0.9\textwidth]{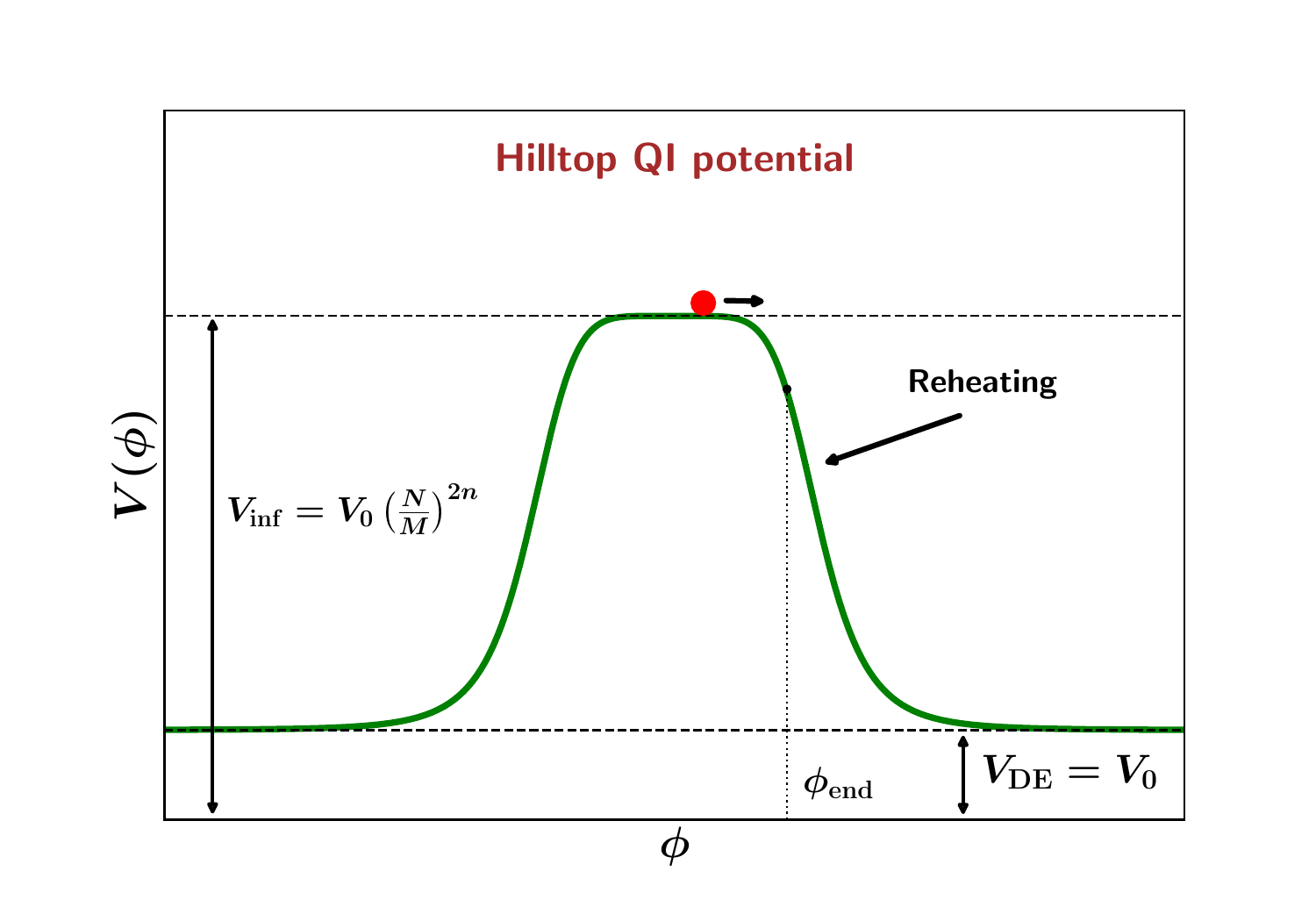}
\caption{The hilltop potential (\ref{eq:hilltop_KKLT}) for quintessential inflation  is schematically illustrated in this figure.}
\label{fig:pot_hilltop_KKLT}
\end{figure}

The second model of quintessential inflation
 to emerge from Eq.~(\ref{eq:plateau_hilltop_KKLT}) is a hilltop with flat wings as shown in Fig.~\ref{fig:pot_hilltop_KKLT}. Indeed, it is easy to show
that the potential 
\beq
V_{\rm hill}(\phi) = V_0 \left\lbrack \frac{N^{2n} + \phi^{2n}}{M^{2n} + \phi^{2n}}\right\rbrack~,  ~~M \ll N
\label{eq:hilltop_KKLT}
\eeq
describes a hilltop with flat wings.
Interestingly, in this case $V(\phi=0)$ corresponds to the {\em maximum} of the potential and not to
its minimum, as was the case for the plateau potential.
This can easily be seen by expanding Eq.~(\ref{eq:hilltop_KKLT}) near $\phi=0$, so that
\beq
V(\phi\simeq 0) \simeq V_0\left (\frac{N}{M}\right )^{2n}\left\lbrace 1 - \left (\frac{1}{M^{2n}} -
\frac{1}{N^{2n}}\right )\phi^{2n}\right\rbrace ~.
\label{eq:mimimum}
\eeq
For $M \ll N$ this expression reduces to
\beq
V(\phi\simeq 0) \simeq V_0\left (\frac{N}{M}\right )^{2n}\left\lbrace 1 - \left (\frac{\phi}{M}\right )^{2n}\right\rbrace ~.
\eeq
Since $V''(\phi=0)<0$ one finds that $V(\phi=0)$ corresponds to the maximum of the hilltop potential
(\ref{eq:hilltop_KKLT}).
We therefore find that, in contrast to the plateau potential,
the flat wings of the hilltop
with $V(\phi \to \pm\infty) + V_0$ are associated with dark energy, while the ground state 
\beq
V_{\rm hill}(\phi = 0) = V_0\left (\frac{N}{M}\right )^{2n} \gg V_0~, ~~M \ll N
\eeq
is responsible for Inflation.

In the hilltop model inflation commences when the field is near the top of its potential at $\phi=0$.
As a result
\beq
V_{\rm inf} \equiv 
V(\phi = 0) = V_0\left (\frac{N}{M}\right )^{2n} \simeq 10^{-10}\,m_p^4~.
\label{eq:hilltop1}
\eeq
From the origin at $\phi=0$ the inflaton rolls down the hilltop towards smaller values of $V$
finally arriving at the flat wing of the potential with $V \simeq V_0$. The value of the potential
at the flat wing gives rise to dark energy so that
\beq
\rho_{\rm DE} \simeq V_0 \simeq 10^{-120}\,m_p^4~.
\label{eq:hilltop2}
\eeq
From Eqs.~(\ref{eq:hilltop1}) \& (\ref{eq:hilltop2}) one gets the following relation between the constants $M$ \& $N$
\beq
\left (\frac{N}{M}\right )^{2n} \sim 10^{110} 
\label{eq:ratio_mM1}
\eeq
Note that Eq.~(\ref{eq:ratio_mM1}) is identical to Eq.~(\ref{eq:ratio_mM})
obtained for the plateau potential !

Since the hilltop potential does not possess a minimum, scalar field oscillations do not occur
and the universe must reheat by a mechanism other than that associated with plateau models
which have a well defined minimum around which the inflaton oscillates. In this connection its important to note that
for $M \ll \phi \ll N$ the potential in Eq.~(\ref{eq:hilltop_KKLT}) has the ultra-steep form
\beq
V(\phi) \sim V_0\left (\frac{M}{\phi}\right )^{2n}~.
\eeq
The exceedingly steep nature of the potential in this region
ensures that $\phi$ rapidly
rolls down $V(\phi)$
towards the asymptotically flat wing with $V= V_0$. Since $\dot{\phi}^2 >> V$ during fast-roll, 
the universe can reheat through the instant
preheating mechanism proposed in \cite{Felder:1998vq,Felder:1999pv} and subsequently applied to models of QI in several papers
including \cite{Sami:2004xk}.
Post-inflationary fast-roll implies $w_\phi \simeq 1$ which results in a blue spectrum for
 the cosmological gravitational wave background as discussed in section \ref{sec:Omega_GW}.

\section{Exponential potentials for QI}
\label{sec:QI_dual_exp}

In the previous section we showed how polynomial potentials could generate 
quintessential inflation. In this section we draw attention to 
a new family of QI models which
can be generated by means of the following
relationship
\beq
V(\phi) = V_0 \,  e^{\pm f(\phi)}~.
\label{eq:QI_exp_pot}
\eeq
Depending upon the form of $f(\phi)$ the potential $V(\phi)$ can be either a plateau or a hilltop.
For instance, if $f_{\rm hill}(\phi)$ has the form of a hilltop, then
hilltop and plateau potentials of QI are determined via
\beq
V_{\rm hill}^{\rm QI}(\phi) = V_0\, e^{f_{\rm hill}(\phi)}~, ~~
V_{\rm pl}^{\rm QI}(\phi) = V_0\, e^{-f_{\rm hill}(\phi)}~. 
\label{eq:QI_exp_relation1}
\eeq
On the other hand, if $f_{\rm pl}(\phi)$ is plateau-like then
\beq
V_{\rm pl}^{\rm QI}(\phi) = V_0\, e^{f_{\rm pl}(\phi)}~, ~~
V_{\rm hill}^{\rm QI}(\phi) = V_0\, e^{-f_{\rm pl}(\phi)}~.
\label{eq:QI_exp_relation2}
\eeq
Note that the potentials $f(\phi)$ on their own can describe inflation \underline{but not DE},
whereas $V(\phi)$ defined by Eq.~(\ref{eq:QI_exp_pot}) successfully describes both inflation and DE.

Below we provide some concrete examples of
the new ansatz given in Eqs.~(\ref{eq:QI_exp_relation1}) and (\ref{eq:QI_exp_relation2}).

\subsection{Exponential based QI potentials from  the hilltop function $f(\phi) = \frac{M^2}{N^2 + \phi^2}$~.}
\label{sec:QI_dual_exp_KKLT}

\begin{itemize}

\item
The potential $V \propto e^{f(\phi)}$  describes a {\em hilltop} model of quintessential inflation
\beq
V(\phi) = V_0\exp\left\lbrack\frac{M^2}{N^2 + \phi^2}\right\rbrack~.
\label{eq:QI_exp_hilltop}
\eeq

In this case the potential at the origin describes Inflation
\beq
V_{\rm inf} \equiv V(\phi \simeq 0) = V_0 \exp\left\lbrack\frac{M^2}{N^2}\right\rbrack \simeq 10^{-10}\mpl^4
\label{eq:QI_exp_hilltop_inflation}
\eeq
while the potential at $\phi \gg M,N$ describes DE
\beq
V_{\rm DE} \equiv V(\phi \to \pm\infty) = V_0 \simeq 10^{-120}\mpl^4~.
\label{eq:QI_exp_hilltop_V0}
\eeq
Consequently one finds
\beq
\frac{M}{N} = \left\lbrack\log{\left (\frac{V_{\rm inf}}{V_{\rm DE}}\right )}\right\rbrack^\frac{1}{2} \simeq 16.
\eeq
Note that in this model the fundamental energy scale $V_0$  in Eq.~(\ref{eq:QI_exp_hilltop_V0})
is small and corresponds to DE, while the large inflationary energy scale
(\ref{eq:QI_exp_hilltop_inflation}) depends upon $V_0,M,N$.

The potential (\ref{eq:QI_exp_hilltop}) can be easily generalised to
\beq
V(\phi) = V_0\exp\left\lbrack\frac{M^{2n}}{N^{2n} + \phi^{2n}}\right\rbrack
\label{eq:QI_exp_hilltop1}
\eeq
which describes a hilltop with flat wings, and leads to
\beq
\frac{M}{N} = \left\lbrack\log{\left (\frac{V_{\rm inf}}{V_{\rm DE}}\right )}\right\rbrack^\frac{1}{2n} ~.
\eeq

\item
The inverse potential $V \propto e^{-f}$ describes a {\em plateau} model of quintessential inflation
\beq
V(\phi) = V_0\exp\left\lbrack-\frac{M^2}{N^2 + \phi^2}\right\rbrack~.
\label{eq:QI_exp_plateau1}
\eeq

For $\phi \gg M,N$ one gets  the asymptotic form
\beq
 V_{\rm inf} \equiv V(\phi \to \pm\infty) \simeq V_0 \simeq 10^{-10}\mpl^4
\label{eq:QI_exp_plateau1a}
\eeq
which describes inflation.

At the origin $\phi = 0$ the potential in Eq.~(\ref{eq:QI_exp_plateau1}) describes DE with
\beq
V_{\rm DE} \equiv V(\phi=0) = V_0\exp\left\lbrack-\frac{M^2}{N^2}\right\rbrack \simeq 10^{-120}\mpl^4
\label{eq:QI_exp_plateau2}
\eeq
Substituting $V_0 \simeq 10^{-10}\mpl^4$ one finds again
\beq
\frac{M}{N} = \left\lbrack\log{\left (\frac{V_{\rm inf}}{V_{\rm DE}}\right )}\right\rbrack^\frac{1}{2} \simeq 16~.
\label{eq:QI_exp_MN}
\eeq
In other words, the constants $M$ and $N$ have comparable values and if
$M \simeq \mpl \sim 10^{18}\, GeV$ then $N \sim 10^{17} - 10^{18}\, GeV$. The QI potential
(\ref{eq:QI_exp_plateau1}) therefore allows one to obtain
\underline{a small value of DE from Planck scale physics.}
Note that in contrast to (\ref{eq:QI_exp_hilltop})
the fundamental energy scale $V_0$ in Eq.~(\ref{eq:QI_exp_plateau1a}) is large and corresponds to Inflation,
while the small DE energy scale
(\ref{eq:QI_exp_plateau2}) 
depends upon $V_0,M,N$.

Eq.~(\ref{eq:QI_exp_plateau1}) can be easily generalized to
\beq
V(\phi) = V_0\exp\left\lbrack-\frac{M^{2n}}{N^{2n} + \phi^{2n}}\right\rbrack
\label{eq:QI_exp_plateau_gen}
\eeq
which once more leads to
\beq
\frac{M}{N} = \left\lbrack\log{\left (\frac{V_{\rm inf}}{V_{\rm DE}}\right )}\right\rbrack^\frac{1}{2n} ~.
\label{eq:QI_exp_MN_gen}
\eeq

\end{itemize}

\subsection{QI from the plateau function $f(\phi) = \frac{(\lambda\phi)^2}{N^2 + \phi^2}$~.}
\label{sec:QI_dual_exp_lambda}

\begin{itemize}
\item The potential $V \propto e^{f}$
 describes a {\em plateau}  model of QI
\beq
V(\phi) = V_0\exp\left\lbrack\frac{(\lambda\phi)^2}{N^2 + \phi^2}\right\rbrack~,
\label{eq:QI_exp_l_plateau}
\eeq
in which the potential at the origin describes DE
\beq
V_{\rm DE} \equiv V(\phi \simeq 0) = V_0 \simeq 10^{-120}\mpl^4~,
\label{eq:QI_exp_l_plateau1}
\eeq
while the potential at $\phi \to \pm\infty$ describes Inflation
\beq
V_{\rm inf} \equiv V(\phi \to \pm\infty) = V_0 e^{\lambda^2} \simeq 10^{-10}\mpl^4~.
\label{eq:QI_exp_l_plateau2}
\eeq
It is easy to show that
\beq
\lambda = \left\lbrack\log{\left (\frac{V_{\rm inf}}{V_{\rm DE}}\right )}\right\rbrack^\frac{1}{2} \simeq 16~,
\eeq
which demonstrates that the tiny value of DE is associated with moderate values of
the parameter $\lambda$.
The potential in Eq.~(\ref{eq:QI_exp_l_plateau}) can be easily generalised to
\beq
V(\phi) = V_0\exp\left\lbrack\frac{({\lambda\phi})^{2n}}{N^{2n} + \phi ^{2n}}\right\rbrack~,
\label{eq:QI_exp_l_plateau_gen}
\eeq
from which follows a simple relation
between the parameters $\lambda$, $V_{\rm inf}$, $V_{_{\rm DE}}$, namely
\beq
\lambda = \left\lbrack \log{\left (\frac{V_{\rm inf}}{V_{_{\rm DE}}}\right )}\right\rbrack^{\f{1}{2n}}~,
\label{eq:QI_exp_lambda1}
\eeq
implying $\lambda \sim O(1)$ for $n \geq 1$.
Note that in this model the fundamental energy scale $V_0$ is small and corresponds to DE.

\item The inverse potential $V \propto e^{-f}$ describes a {\em hilltop}  model of Quintessential Inflation
\beq
V(\phi) = V_0\exp\left\lbrack-\frac{(\lambda\phi)^2}{N^2 + \phi^2}\right\rbrack~.
\label{eq:QI_exp_l_hilltop}
\eeq
Note that this potential describes a Gaussian with asymptotically flat wings.
In this case the potential at the origin describes Inflation
\beq
V_{\rm inf} \equiv V(\phi \simeq 0) = V_0 \simeq 10^{-10}\mpl^4~,
\label{eq:model2}
\eeq
while the potential at $\phi \to \pm\infty$ describes DE
\beq
V_{\rm DE} \equiv V(\phi \to \pm\infty) = 
V_0 e^{-\lambda^2} \simeq 10^{-120}\mpl^4~.
\label{eq:model2a}
\eeq
Eqs.~(\ref{eq:model2}) and (\ref{eq:model2a}) imply that
\beq
\lambda = \left\lbrack\log{\left (\frac{V_{\rm inf}}{V_{\rm DE}}\right )}\right\rbrack^\frac{1}{2} \simeq 16~.
\label{eq:QI_exp_l_hilltop_lambda}
\eeq
It is interesting to note that the value of the constant $\lambda \simeq 16$ is of the same order of magnitude
as $\alpha^{-1/2} \simeq 12$, where $\alpha = 1/137$ is the fine structure constant.
The QI potential (\ref{eq:QI_exp_l_hilltop}) therefore succeeds in deriving
a microscopically small value of DE from reasonably valued constants in the theory.

The potential in Eq.~(\ref{eq:QI_exp_l_hilltop}) can be easily generalised to
\beq
V(\phi) = V_0\exp\left\lbrack-\frac{({\lambda\phi})^{2n}}{N^{2n} + \phi ^{2n}}\right\rbrack~,
\label{eq:QI_exp_l_hilltop_gen}
\eeq
where $V(\phi)$ is symmetric and consists of three distinct
regions:

\begin{enumerate}

\item For $|\phi| \ll N/\lambda$

\beq
V(\phi) \simeq V_0\left\lbrack 1 - \left (\frac{\lambda\phi}{N}\right )^{2n}\right\rbrack~.
\label{eq:pot1_asym1}
\eeq
Thus, for small value of $|\phi|$ the potential (\ref{eq:QI_exp_l_hilltop_gen}) resembles a hilltop.

\item In the intermediate region $N/\lambda \ll |\phi| \ll N$ the potential (\ref{eq:QI_exp_l_hilltop_gen})
decreases rapidly

\beq
V(\phi) \simeq V_0\, e^{^{-\left (\frac{\lambda\phi}{N}\right )^{2n}}}~.
\label{eq:pot1_asym2}
\eeq
The rapid decrease in the value of $V(\phi)$ ensures that the
universe can reheat through the mechanism of instant preheating \cite{Felder:1998vq,Felder:1999pv}; also see
\cite{Geng:2017mic}.

\item As $\phi(t)$ rolls to still larger values, $|\phi| \gg N$, the potential
(\ref{eq:QI_exp_l_hilltop_gen}) approaches its asymptotically flat form
\beq
V_{_{\rm DE}} = V_0\,e^{^{-\lambda^{2n}}}~.
\label{eq:pot1_asym3}
\eeq
At this late epoch $V(\phi)$ resembles a cosmological constant which results in
late time acceleration once $V_{_{\rm DE}}$ begins to dominate
over the matter density. 
As a result late-time expansion in our model mimicks $\Lambda$CDM 
\beq
a(t) \propto \left (\sinh{\frac{3}{2}\sqrt{\frac{\Lambda}{3}}ct}\right )^{2/3}
\eeq
where $\frac{\Lambda}{8\pi G} \equiv V_{_{\rm DE}}$.

Finally note that Eq.~(\ref{eq:QI_exp_l_hilltop_lambda}) generalises to
\beq
\lambda = \left\lbrack\log{\left (\frac{V_{\rm inf}}{V_{\rm DE}}\right )}\right\rbrack^\frac{1}{2n}~, 
\eeq
which again implies that $\lambda \sim O(1)$.
Note that in this model the energy scale $V_0$ is large and corresponds to Inflation.

\end{enumerate}

\end{itemize}

It is interesting to note that the pair of potentials  discussed above  are actually equivalent 
(see appendix \ref{sec:app_eqv}) since one can be derived from the other by the parameter redefinition 
$\f{M}{N} \longrightarrow \lambda$.  Hence, later in this paper, we will discuss the  inflationary dynamics  and the
associated relic GW background for only one of the two models.

\bigskip

Finally note that models of QI very similar to those discussed above can be constructed
using the cosine hyperbolic potential which has earlier been discussed in connection with tracker dark matter in \cite{Sahni:1999qe}
and tracker dark energy in \cite{Bag:2017vjp}.

\begin{enumerate}

\item
It is easy to see, for instance, that
\beq
V_{\rm pl}(\phi) = V_0\,\cosh{\left\lbrack\frac{\lambda\phi}{\sqrt{N^2 + \phi^2}}\right\rbrack}
\eeq
describes a plateau model of QI, while its inverse
\beq
V_{\rm hill}(\phi) = V_0\,\cosh{\left\lbrack\frac{-\lambda\phi}{\sqrt{N^2 + \phi^2}}\right\rbrack}
\eeq
describes hilltop QI.

\item Similarly
\beq
V(\phi) = V_0\cosh{\left\lbrack\frac{M}{\sqrt{N^2 + \phi^2}}\right\rbrack}~.
\eeq
describes hilltop QI, whereas its inverse
\beq
V(\phi) = V_0\cosh{\left\lbrack\frac{-M}{\sqrt{N^2 + \phi^2}}\right\rbrack}~.
\eeq
describes plateau QI.

\end{enumerate}

\section{Quintessential Inflation from asymptotically non-flat potentials}
\label{sec:QI_nonflat}

\subsection{Chaotic QI}
\label{sec:QI_dual_nonflat}

The potentials discussed in the previous sections were asymptotically flat since $V \to {\rm constant}$ at
$\phi \to \pm \infty$. In this section we discuss simple generalisations of Eqs.~(\ref{eq:plateau_hilltop}) \& 
(\ref{eq:plateau_hilltop_KKLT}) which give rise to potentials which are not
asymptotically flat. 

\begin{itemize}

\item
Consider for instance 
\beq
V(\phi) = V_0\,v(\varphi) ~~~{\rm and} ~~ V(\phi) = V_0\,\left[v(\varphi)\right]^{-1}~,
\label{eq:nonflat1}
\eeq
with
\beq
v(\varphi) = \left\lbrack \frac{A + \varphi^{2n+\alpha}}{B + \varphi^{2n}}\right\rbrack~, 
\label{eq:nonflat2}
\eeq
where $A,\, B$ are constants, $\varphi = \phi/m_p$, ~$0 < \alpha < 1$ and $n = 0,1,2,3...$.
This potential has the following properties
\ber
V(\varphi = 0) &=& V_0\,\left (\frac{A}{B}\right ) \, ; \nonumber\\
V(\varphi \to \pm \infty) &\simeq& V_0\, \varphi^\alpha ~. 
\label{eq:nonflat3}
\eer
Thus, for $A \ll B$, the potential at the origin can describe DE. 
In the context of inflation one might note that in marked contrast to plateau potentials, $V(\phi)$ in Eq.~(\ref{eq:nonflat3}) grows as $|\phi|$ increases. This interesting property leads to {\em chaotic inflation}
which allows inflation to commence
from Planck scale values with equipartition initial conditions, namely: 
\beq
{\dot\phi_i}^2 \sim V(\phi_i) \simeq m_p^4~.
\eeq
 Monomial potentials with $\alpha \leq 4/3$  have been extensively
discussed in the context of monodromy inflation in \cite{Silverstein:2008sg,McAllister:2008hb}.  Unfortunately, they are in tension\footnote{Monomial potentials with $\alpha < 1$ can become compatible with the CMB data if the reheating history of the universe results in a higher number of e-folds of inflation in between the Hubble-exit of the CMB  pivot scale and the end of inflation, namely, $N_* > 60$. } with the latest CMB observations \cite{Planck_inflation,BICEP:2021xfz,Mishra:2022ijb}.
The inverse of Eq.~(\ref{eq:nonflat2}) namely $V(\phi) = V_0\,\left[v(\varphi)\right]^{-1}$ 
describes hilltop QI with 
\ber
V(\varphi = 0) &=& V_0\,\left (\frac{B}{A}\right )~, ~~ B \gg A \nonumber\\
V(\varphi \to \pm \infty) &\simeq & {V_0}/{\varphi^{\alpha}}~.
\eer
One immediately finds that, in contrast to the hilltop potentials discussed earlier, this particular hilltop
{\em does not have flat wings}. Instead the potential slopes gently downwards towards the asymptotic form
$V \propto \phi^{-\alpha}$ typical of the Ratra-Peebles potential \cite{Ratra:1987rm}.
Thus this hilltop model unifies inflation with dynamical dark energy, as in the original QI construct \cite{Peebles:1998qn}.

\n
Note that for $n=0$ the potential in Eq.~(\ref{eq:nonflat2}) acquires the following simple form:
\begin{enumerate}

\item The potential 
\beq
V(\phi) = V_0\left\lbrack \frac{A + \varphi^\alpha}{C}\right\rbrack~, ~~ C = 1+B
\eeq
is shaped like a well, with the ground state at $V(0)$ 
describing dark energy.

\item The inverse potential 
\beq
V(\phi) = V_0 \left\lbrack \frac{C}{A + \varphi^\alpha}\right\rbrack
\eeq
describes a hilltop with the potential at $V(0)$ giving rise to inflation.

\end{enumerate}
\n

\begin{figure}[htb]
\centering
\includegraphics[width=0.9\textwidth]{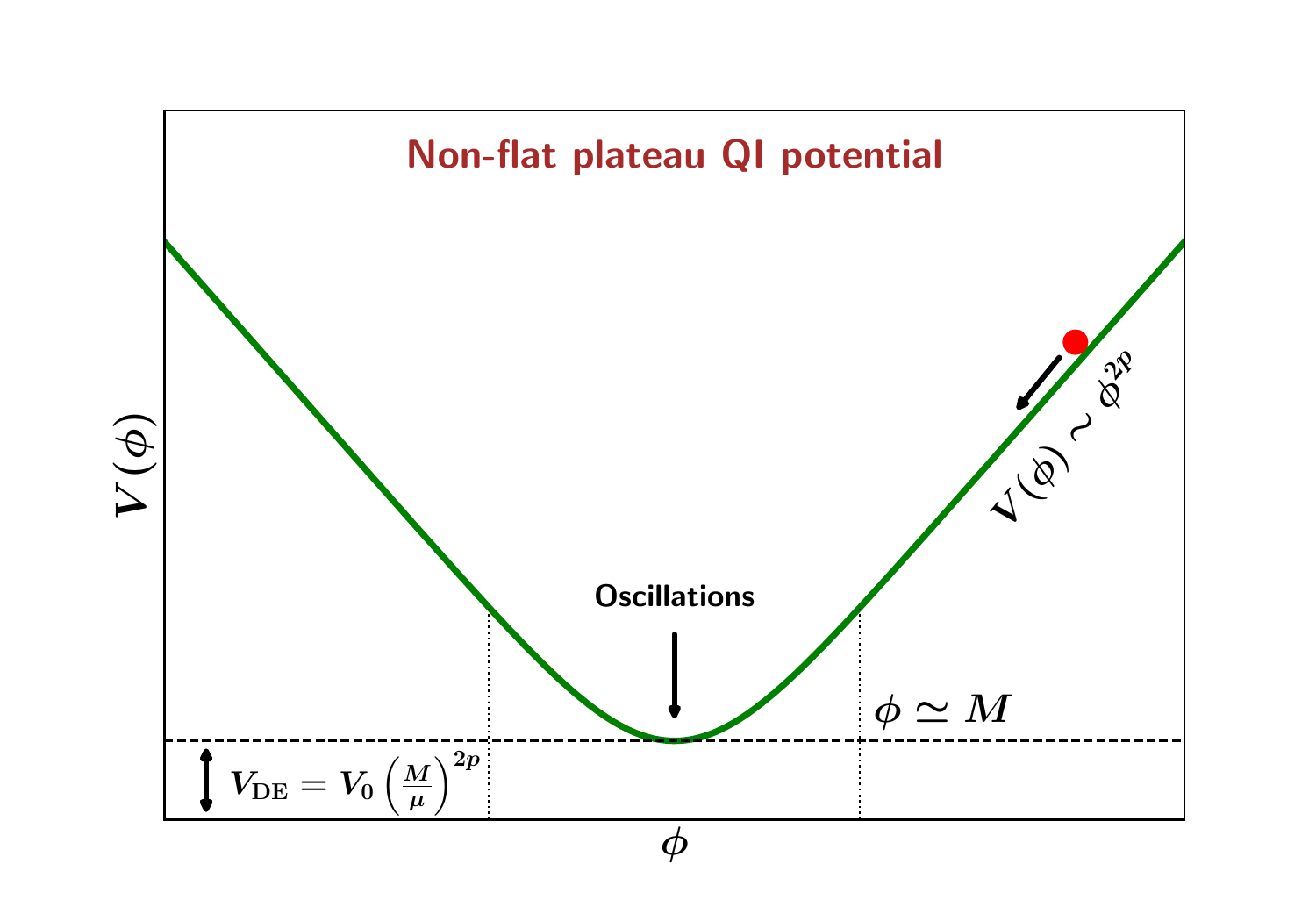}
\caption{The non-flat QI  potential (\ref{eq:pot_QI_nf_tanh})  is schematically shown in this figure.}
\label{fig:pot_QI_tanh_plateau}
\end{figure}

\item
Another interesting family of potentials also gives similar results.

Consider for instance the potential
\beq
V(\phi) = V_0 \frac{\left ({\phi}/{\mu}\right)^{2p}}{\tanh^{2p}
\left (\frac{\phi}{M}\right)}
\label{eq:pot_QI_nf_tanh}
\eeq
Since $\tanh{(x)} \simeq x$, for $x \ll 1$ and 
$\tanh{(x)} \simeq 1$, for $x \gg 1$, one finds
\ber
V(\phi) &\simeq& V_0 \left (\frac{\phi}{\mu}\right)^{2p},
~~{\rm when} ~\phi \gg M ~~{\rm and} \nonumber\\
V(\phi) &\simeq& V_0\left (\frac{M}{\mu}\right )^{2p}, ~~{\rm when} ~\phi \ll M.
\eer
This model describes monodromy inflation when $\phi \gg M, ~p \leq 1/2$ and a cosmological
constant (DE) at $\phi \simeq 0$. Oscillations of the field
 around $\phi \simeq 0$
give rise to preheating. Note that $M \ll \mu$ is required to give
acceptable values of DE at present.

\begin{figure}[htb]
\centering
\includegraphics[width=0.9\textwidth]{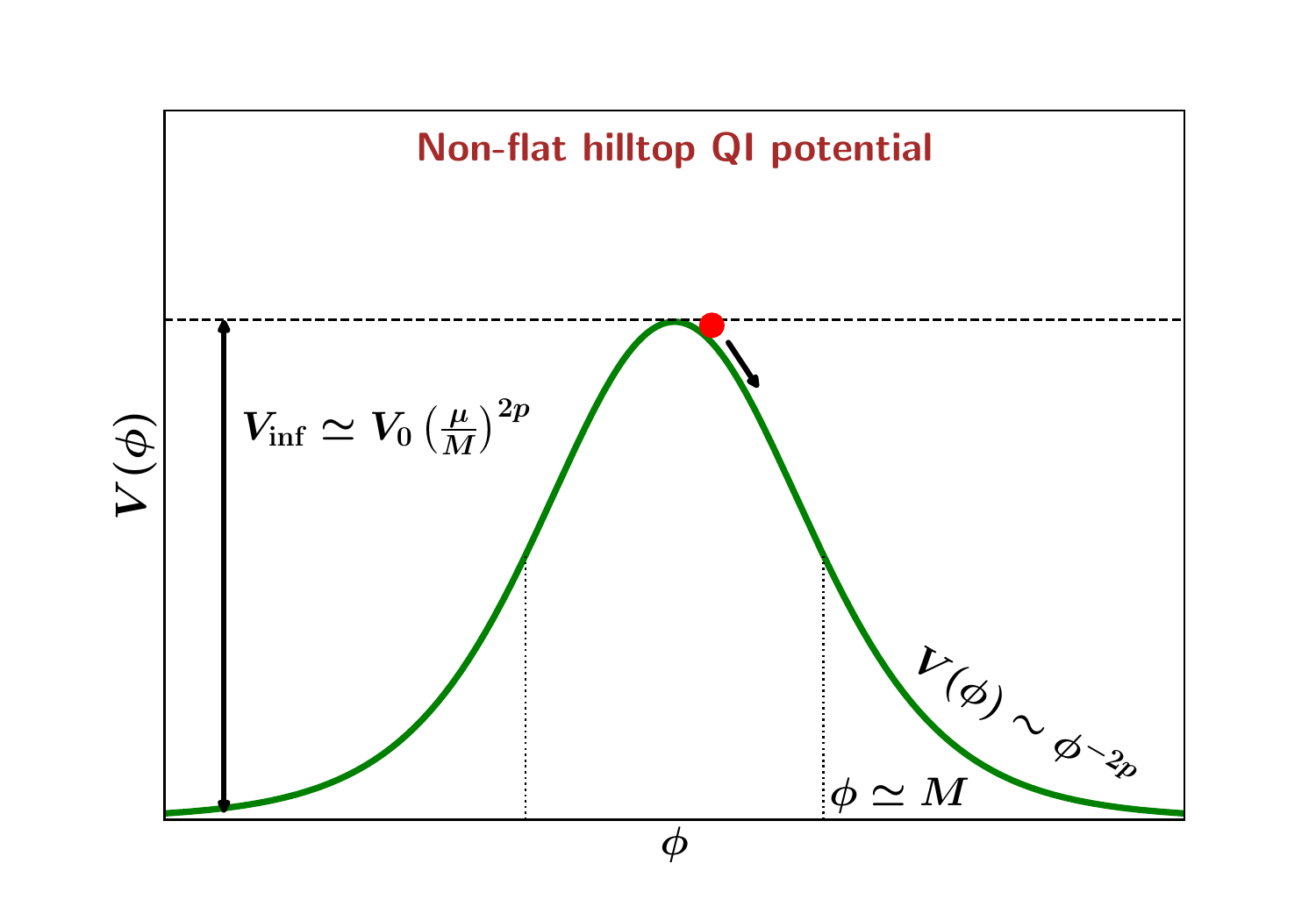}
\caption{The non-flat QI hilltop potential (\ref{eq:pot_QI_nf_tanh_hilltop})  is schematically illustrated in this figure.}
\label{fig:pot_QI_tanh_hilltop}
\end{figure}

The inverse potential, namely
\beq
V(\phi) = V_0 \frac{\tanh^{2p}
\left (\frac{\phi}{M}\right)}{\left ({\phi}/{\mu}\right)^{2p}}
\label{eq:pot_QI_nf_tanh_hilltop}
\eeq
describes hilltop QI with
\ber
V(\phi) &\simeq& V_0\left (\frac{\mu}{M}\right )^{2p}, ~~{\rm when} ~\phi \ll M
~~{\rm and} \nonumber\\
V(\phi) &\simeq& V_0 \left (\frac{\mu}{\phi}\right )^{2p}, ~~{\rm for
} ~\phi \gg M
\eer
where $M \ll \mu$ and
 small values of the parameter $p \ll 1$ provide good agreement
with observations.

The non-flat plateau and hilltop potentials for QI discussed above have been schematically illustrated in Figs.~\ref{fig:pot_QI_tanh_plateau} and \ref{fig:pot_QI_tanh_hilltop} respectively.

\end{itemize}

\subsection{A logarithmic potential for QI}
\label{sec:QI_log}

Another  asymptotically non-flat QI potential is given by
\beq
V(\phi) = V_0\log{\left\lbrace\frac{m^2+\phi^2}{m_p^2}\right\rbrace}~.
\label{eq:QI_log}
\eeq
One notes that this potential has plateau-like features for moderate values of $|\phi|$ and becomes
unbounded as $|\phi| \to \infty$.
Rewriting Eq.~(\ref{eq:QI_log}) as
\beq
V(\phi) = V_0\log{\left (\frac{m^2}{m_p^2}\right )} +
V_0\log{\left\lbrack 1 + \frac{\phi^2}{m^2}\right\rbrack}
\label{eq:QI_log1}
\eeq
we find that the second term in the RHS vanishes at $\phi=0$.
Therefore at late times, after scalar field oscillations have declined and the field has settled down to its
ground state, DE is described by the first term in the RHS, namely
\beq
V(\phi=0) \equiv V_{\rm DE} = V_0\log{\left (\frac{m^2}{m_p^2}\right )}~.
\label{eq:QI_log_DE}
\eeq

\begin{figure}[htb]
\centering
\includegraphics[width=0.9\textwidth]{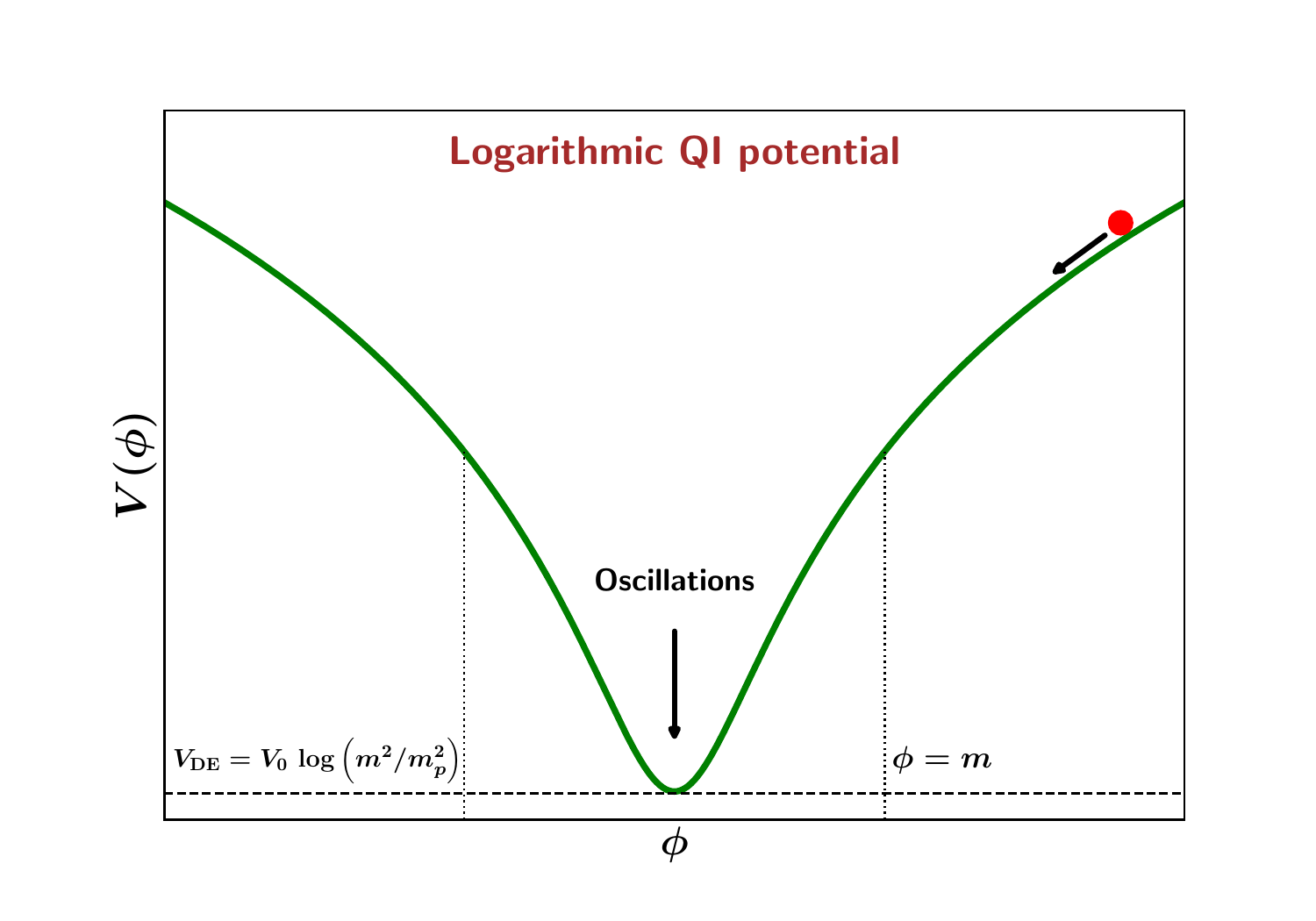}
\caption{The  logarithmic  potential for QI (\ref{eq:QI_log})  is schematically shown in this figure.}
\label{fig:pot_QI_log}
\end{figure}

Since $V_0 \sim 10^{-10}m_p^4$ is suggested by CMB constraints on inflation \footnote{Inflation occurs in
this model when
$|\phi| \gg m$.} one finds from Eq.~(\ref{eq:QI_log_DE})
 that the value of the mass scale $m$ must be very close to $m_p$ in order that
$V_{\rm DE} \sim 10^{-120}\,m_p^4$.

Post-inflation, when $|\phi| \ll m$, the second term in Eq.~(\ref{eq:QI_log1}) reduces to
\beq
V_0\log{\left\lbrack 1 + \frac{\phi^2}{m^2}\right\rbrack} \simeq \left (\frac{V_0}{m^2}\right ){\phi^2}~,
\eeq
which implies that the scalar field settles to its ground state at $\phi = 0$ via small oscillations.
Consequently reheating in this scenario is likely to proceed along lines similar to that in other
plateau-like inflationary potentials.
Logarithmic potentials are known to occur in supersymmetric theories and our phenomenological potential (\ref{eq:QI_log}) is inspired by the Witten-ORaifeartaigh potential discussed in \cite{Witten:1981kv,ORaifeartaigh:1975nky,Martin:2013tda}. The CMB observables $\lbrace n_{_S},r \rbrace$ for this potential are given in tab.~\ref{table:1}.

\begin{table}[h]
\begin{center}
 \begin{tabular}{||c|c|c|c|c||} 
 \hline\Tstrut
 $N_e$ &  $n_{_S}$ & $r$ \\ [1ex] 
 \hline\hline\Tstrut
 60 & 0.97751 & 0.0311 \\ [1.2ex] 
 \hline\Tstrut
 50 & 0.97281 &  0.0252 \\ [1.2ex]
 \hline
\end{tabular}
\captionsetup{
	justification=raggedright,
	singlelinecheck=false
}
\caption{The inflationary observables $\lbrace n_{_S},r \rbrace$ are listed for the logarithm potential (\ref{eq:QI_log}) of quintessential inflation which are compatible with the latest  CMB  2$\sigma$ constraints of Planck 2018 and BICEP/Keck observations \cite{Planck_inflation, Mishra:2022ijb}, namely $\ns \in \l[ 0.957, \, 0.976 \r]$, $r\leq 0.036$.}
\label{table:1}
\end{center}
\end{table}

\section{Inflationary dynamics of plateau and hilltop potentials}
\label{sec:QI_inf_dual}

The following set of equations describes the cosmological expansion of a flat Friedman-Robertson-Walker (FRW) 
universe sourced by a minimally coupled scalar field
\ber
H^2 = \frac{1}{3m_p^2} \, \rho_{\phi}=\frac{1}{3m_p^2} \l[\frac{1}{2}{\dot\phi}^2 +V(\phi)\r],
\label{eq:friedmann1}\\
\dot{H}=\frac{\ddot{a}}{a}-H^2=-\frac{1}{2 m_p^2}\,\dot{\phi}^2,
\label{eq:friedmann2}\\
{\ddot \phi}+ 3\, H {\dot \phi} + V'(\phi) = 0~.
\label{eq:phi_EOM}
\eer
If cosmic expansion is inflationary then the 
duration of inflation can be described by the total number of e-foldings 
\beq
\Delta N_e = N_e^i - N_e^{\rm end} = \log_e{\frac{a_{\rm end}}{a_i}}=\int_{t_i}^{t_{\rm end}} H(t) dt,
\label{eq:efolds}
\eeq
where $N_e$ denotes the number of e-foldings {\em before the end of inflation}: 
with $N_e=N_e^i$ corresponding to the beginning of inflation and
$N_e = N_e^{\rm end} = 0$ corresponding to the end of inflation.
$a_i$ and $a_{\rm end}$ denote the scale factor at the beginning and
end of inflation respectively. 
We denote $N_*$ as the number of e-foldings (before the end of inflation) when the CMB pivot scale $k_*=(aH)_*=0.05~\mpc$ left the comoving Hubble
radius during inflation. We initially choose  $N_*=60$ although the precise value of
$N_*$ depends upon the reheating temperature.
Quasi-de Sitter like expansion ensues when the inflaton field rolls slowly down the
potential $V(\phi)$. This slow-roll phase of inflation,
 is characterised by the first two Hubble slow-roll parameters $\epsilon_H$, $\eta_H$ \cite{Baumann_TASI}
\ber
\epsilon_H = -\frac{\dot{H}}{H^2}=\frac{1}{2 m_p^2}\frac{\dot{\phi}^2}{H^2},
\label{eq:epsilon_H}\\
\eta_H = -\frac{\ddot{\phi}}{H\dot{\phi}}=\epsilon_H+\frac{1}{2\epsilon_H}\frac{d\epsilon_H}{dN_e}~,
\label{eq:eta_H}
\eer
where 
$\epsilon_H,~\eta_H\ll 1~,$
during the slow-roll regime. During slow-roll scalar field
perturbations  are  usually quantified in terms of  the comoving curvature perturbation
$\zeta$ and its  power spectrum \cite{Baumann_TASI,Baumann:2018muz}
\beq
{\cal P}_{\zeta} = \frac{1}{8\, \pi^2}\l(\frac{H}{m_p}\r)^2 \f{1}{ \epsilon_H}~.
\label{eq:Ps_slow-roll}
\eeq
On large scales the power spectrum typically takes the form of a power law 
\beq
{\cal P}_{\zeta}(k)=A_{_S}\l(\frac{k}{k_*}\r)^{\ns-1},
\label{eq:PS_power-law}
\eeq
where $A_{_S}={\cal P}_{\zeta}(k_*)$ is the amplitude of the scalar power spectrum at the pivot scale. The scalar spectral tilt $\ns$ and the tensor to scalar ratio $r$, in the slow-roll regime,
are given by \cite{Baumann_TASI}
\ber
\ns &=& 1+2\eta_H-4\epsilon_H~,
\label{eq:ns}\\
r &=& 16~\epsilon_H~,
\label{eq:r}
\eer
 with recent CMB observations \cite{Planck_inflation,BICEP:2021xfz,Mishra:2022ijb} implying
\beq
\ns \in \l[0.957,\, 0.976\r]~, \quad r(k_*) \leq 0.036 ~~ \mbox{at} ~~ 95\% ~ \mbox{C.L}
\label{eq:CMB_ns_r_constraint}
\eeq
and
\beq
A_{_S} \equiv {\cal P}_{\zeta}(k_*) = 2.1\times 10^{-9}
\label{eq:CMB_power_constraint}
\eeq
where $k_*=(aH)_*=0.05~\mpc$ marks the pivot scale. It is interesting to note that the relatively low upper bound on $r$ 
can easily be accommodated in our models of quintessential inflation.

\subsection{Inflationary dynamics of the KKLT type QI models}
\label{sec:QI_inf_dual_KKLT}

Consider first the
KKLT model \cite{Kachru:2003aw} to which our potential (\ref{eq:plateau_KKLT}) reduces when $M=0$, namely
\beq
V(\phi)=V_0\frac{\phi^{2n}}{\phi^{2n}+N^{2n}}~,
\label{eq:base_KKLT}
\eeq
where $V_0$ fixes the overall CMB normalization given by eqs.~(\ref{eq:Ps_slow-roll}) \& (\ref{eq:CMB_power_constraint}).

\begin{figure}[htb]
\centering
\includegraphics[width=0.8\textwidth]{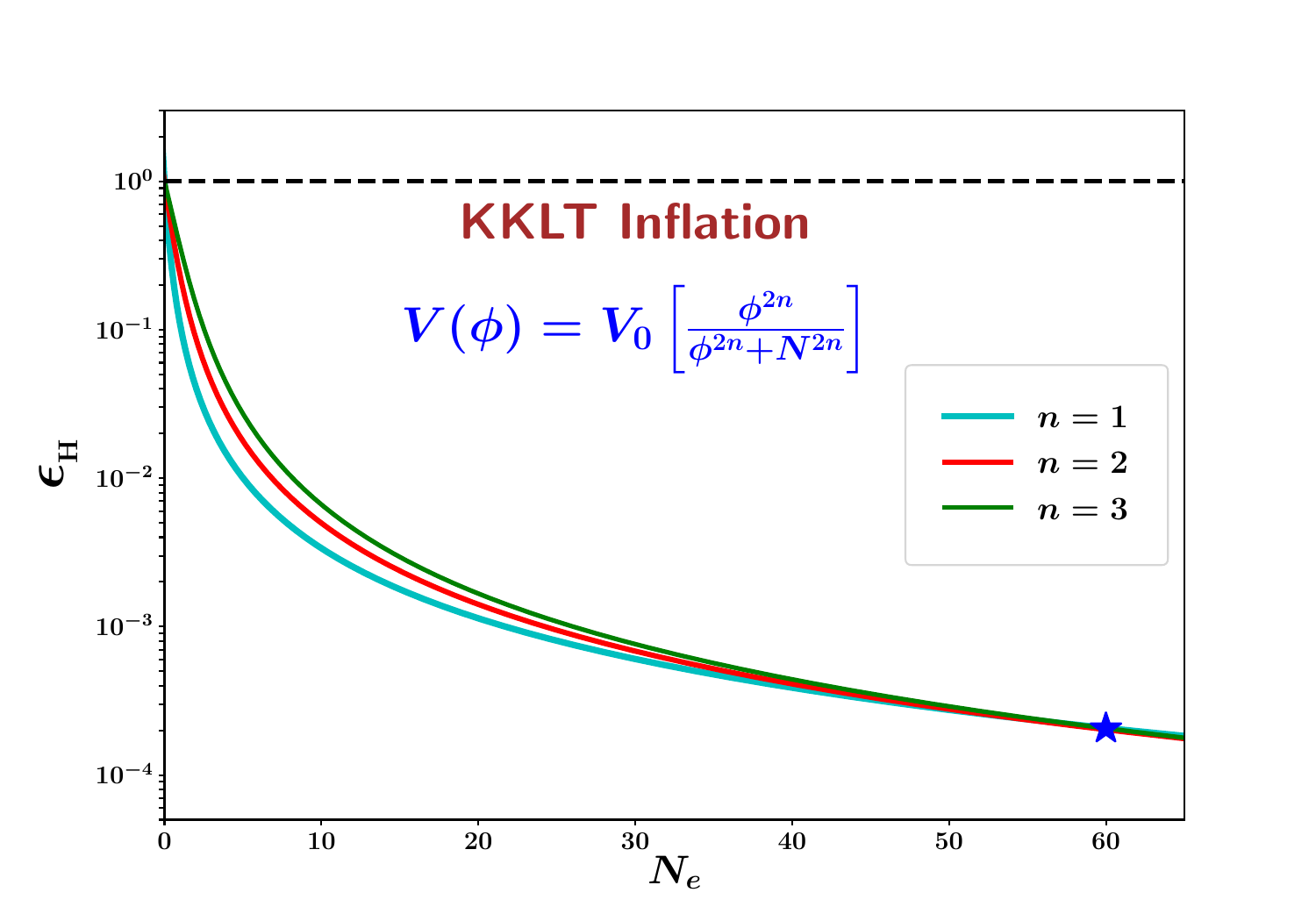}
\caption{The  first slow-roll parameter $\epsilon_H$ 
is plotted against  the number of inflationary
e-folds $N_e$ before the end of KKLT inflation (\ref{eq:base_KKLT}) for $n=1,\,  2, \,  3,$ in cyan, red and green colours respectively.
}
\label{fig:KKLT_epsilon}
\end{figure}

\begin{figure}[htb]
\centering
\includegraphics[width=0.8\textwidth]{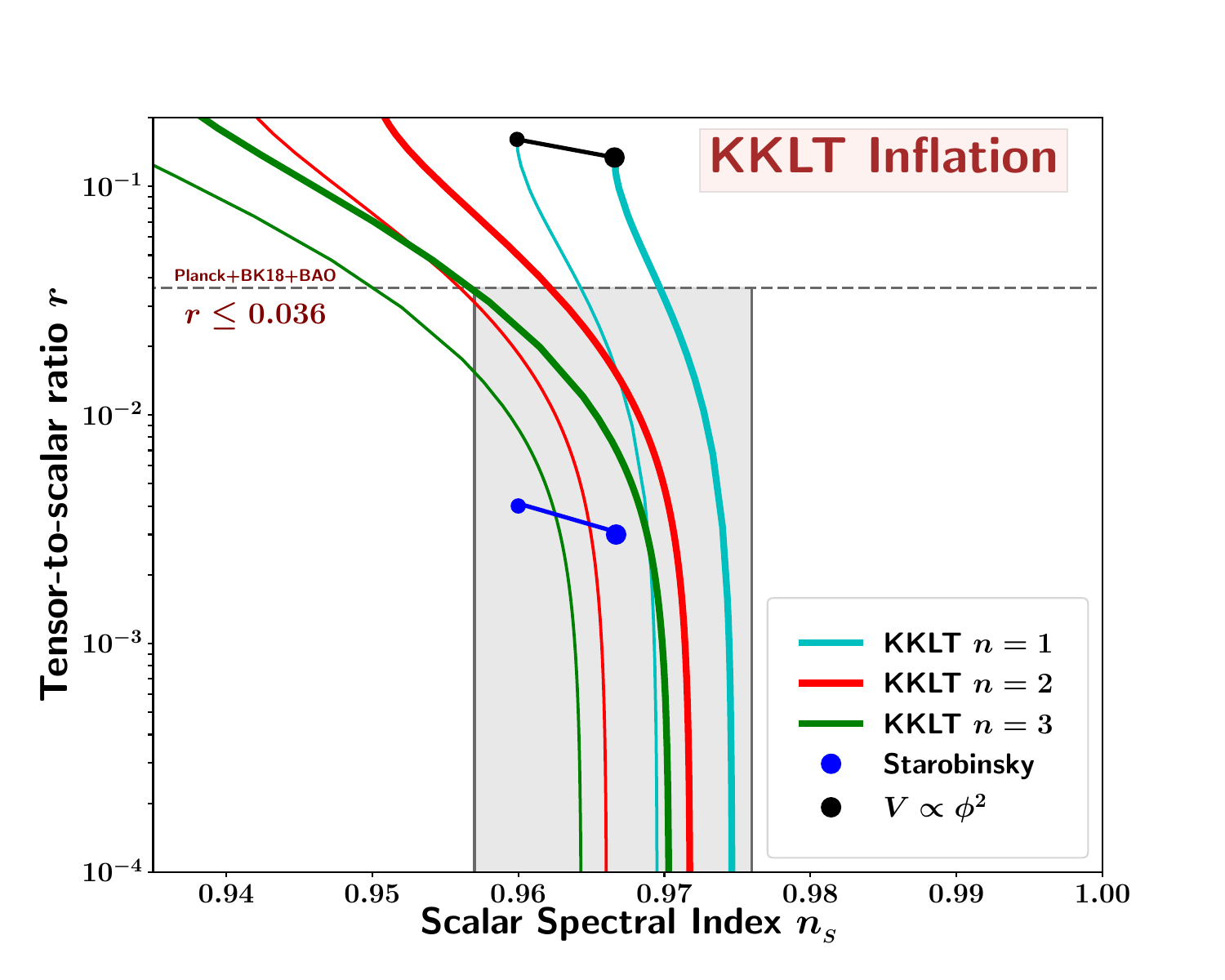}
\caption{The tensor-to-scalar ratio $r$ and scalar spectral tilt $n_{_S}$ are plotted for the KKLT inflationary potential (\ref{eq:base_KKLT}) with $n=1, \, 2, \, 3,$ shown in cyan, red and green respectively (thicker lines correspond to $N_e = 60$ while thinner lines correspond to $N_e = 50$, as per  the standard convention). The CMB 2$\sigma$ bound $0.957 \leq n_{_S} \leq 0.976$
 and the upper bound on  $r$, given in  Eq.~(\ref{eq:CMB_ns_r_constraint}), are
 indicated by the grey colour shaded region. Note that KKLT inflation also satisfies the CMB constraints for the case of $n=4$, however which we do not show the corresponding $r$  vs $n_{_S}$ plot in order to avoid clutter. The flow lines reflect different values of the parameter $N$ in the potential (\ref{eq:base_KKLT}) for a given value of $n$.}
\label{fig:KKLT_r_ns}
\end{figure}

An important physical quantity is the value of the Hubble parameter during inflation which can easily be
deduced by rewriting Eq.~(\ref{eq:Ps_slow-roll}) as
\beq
\frac{H}{m_p} = \sqrt{8\pi^2 \, \epsilon_H \, {\cal P}_\zeta }
\label{eq:Hubble1}
\eeq
As shown in Fig.~\ref{fig:KKLT_epsilon}
 the slow roll parameter $\epsilon_H$ depends upon the number of inflationary
e-foldings, $N_e$, before the end of inflation. One should note that the value of $\epsilon_H$ 
{\em increases} as $N_e$ decreases. 
Substituting $\epsilon_H \simeq 10^{-4}$ and ${\cal P}_\zeta \simeq 2.1\times 10^{-9}$ one finds\,\footnote{The value
of $\epsilon_H$ depends upon the number of inflationary e-foldings, $N_e$, which is sensitive to the
value of the reheating temperature. $\epsilon_H \simeq 10^{-4}$ corresponds to about 60 e-folds before
inflation ends; see Fig.~\ref{fig:KKLT_epsilon}.}
\beq
\frac{H}{m_p} \simeq 4\times 10^{-6}~.
\label{eq:Hubble2}
\eeq
As discussed later in section \ref{sec:GW}, 
the inflationary Hubble parameter plays a key role in determining the amplitude of
relic gravitational waves which are created during inflation \cite{Starobinsky:1979ty}. Figure \ref{fig:KKLT_r_ns} demonstrates that the KKLT potential (\ref{eq:base_KKLT}) satisfies the CMB constraints for $n=1, \, 2, \, 3, \, 4$. Hence our plateau QI potential (\ref{eq:plateau_KKLT})  for $M\ll N$ also satisfies the CMB constraints.

\bigskip

  However the hilltop QI potential (\ref{eq:hilltop_KKLT}) discussed in section \ref{sec:QI_hilltop_KKLT} does not satisfy the CMB constraints of Planck 2018 because of its small scalar spectral index $n_{_S} < 0.948$. Although the tensor-to-scalar ratio $r$ for this potential is extremely small \textit{i.e} $r \ll 10^{-3}$. Moreover,   given the fact that inflation is possible for $\phi \leq M \ll N$ and since  $M \ll m_p$, inflation happens for a small range of values of  $\phi$ near the top of the potential, namely $\phi \in [-M,M]$. This leads to the problem of initial conditions for inflation for the plateau QI potential  (\ref{eq:hilltop_KKLT}). This 
appears to be a  problem for most small-field hilltop models discussed in the literature
and we refrain from  discussing this model any further in this paper. 

\subsection{Inflationary dynamics of exponential QI models}
\label{sec:QI_inf_dual_exp_KKLT}

\begin{large}
\textbf{(A) Hilltop QI:}
\end{large}
\medskip

\begin{figure}[htb]
\centering
\includegraphics[width=0.8\textwidth]{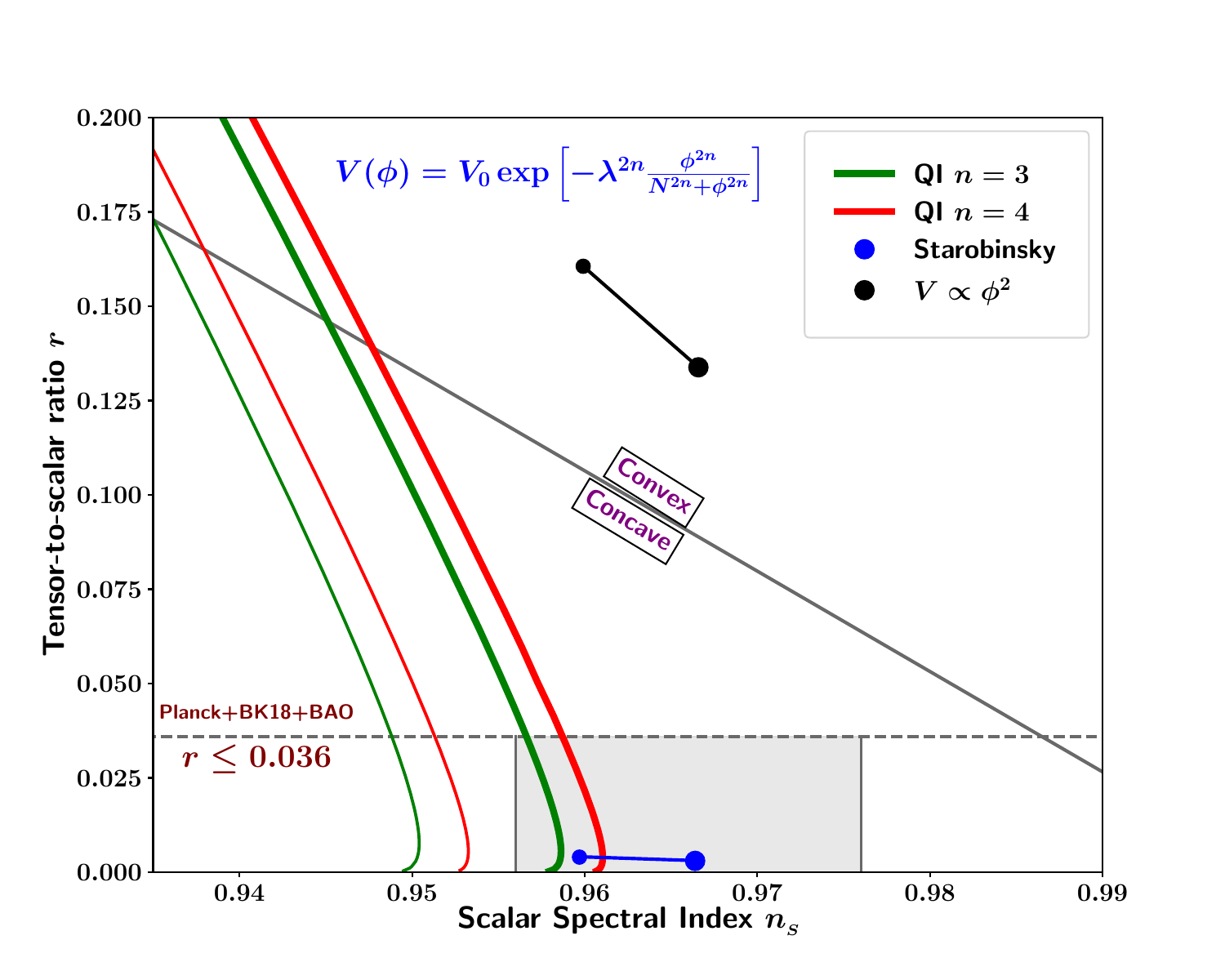}
\caption{The tensor-to-scalar ratio $r$ and scalar spectral tilt $n_{_S}$ are plotted for the hilltop QI potential (\ref{eq:QI_exp_l_hilltop_genA}) for $n=3,\, 4,$  plotted in green and red colours respectively
(thicker lines correspond to $N_e = 60$ while thinner lines correspond to $N_e = 50$, as per  the standard convention). The CMB 2$\sigma$ bound $0.957 \leq n_{_S} \leq 0.976$
 and the upper bound on  $r$, given in  Eq.~(\ref{eq:CMB_ns_r_constraint}), are
 indicated by the grey colour shaded region.
The flow lines reflect different values of the parameter $N$ for a given $n$.
}
\label{fig:QI_exp_minus_l_r_ns}
\end{figure}

As shown in section \ref{sec:QI_dual_exp}, hilltop QI can also be generated by the potential 
(\ref{eq:QI_exp_l_hilltop_gen}) namely
\beq
V(\phi) = V_0\exp\left\lbrack-\frac{(\lambda\phi)^{2n}}{N^{2n} + \phi^{2n}}\right\rbrack~.
\label{eq:QI_exp_l_hilltop_genA}
\eeq
Predictions for the CMB observables $\lbrace n_{_S},r \rbrace$ are
 shown in Fig.~\ref{fig:QI_exp_minus_l_r_ns}. Note that this potential satisfies the 
CMB 2$\sigma$ constraints for $n=3,\,4$ (but not for $n=1,\,2$).

\begin{figure}[t]
\begin{center}
\includegraphics[width=0.495\textwidth]{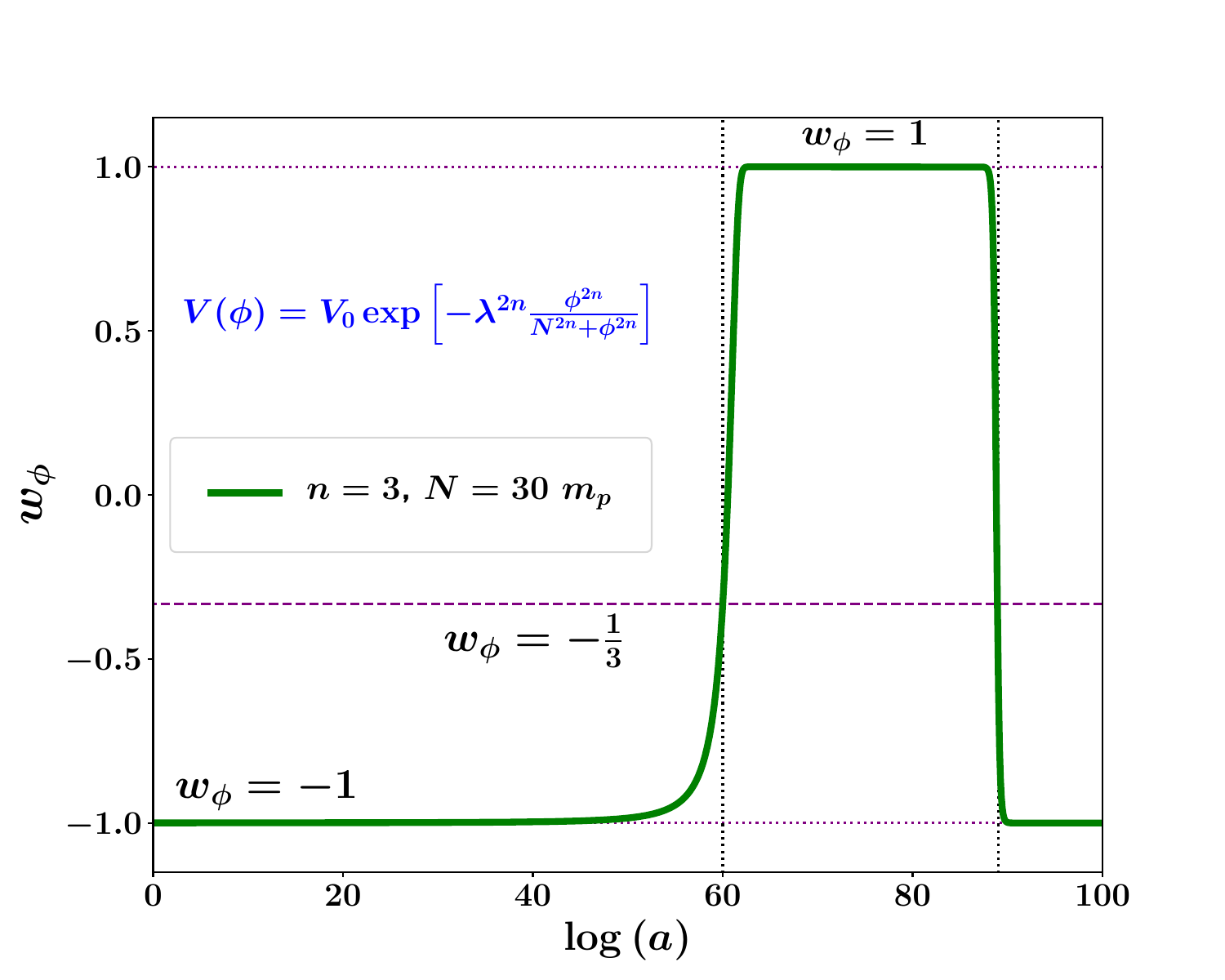}
\includegraphics[width=0.495\textwidth]{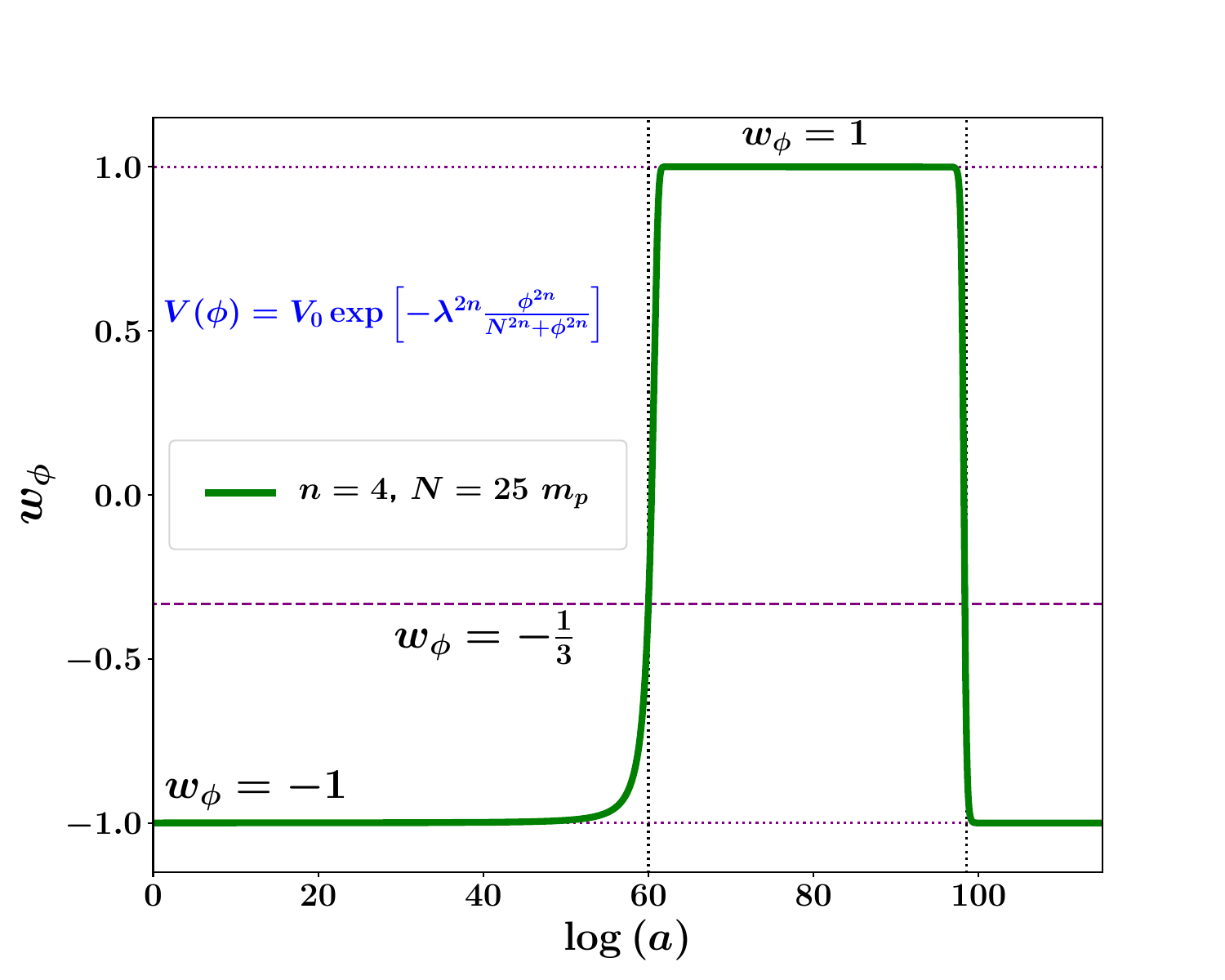}
\caption{Evolution of the  EOS of the inflaton field is plotted for the hilltop QI potential 
(\ref{eq:QI_exp_l_hilltop_genA}) in the {\bf left panel} for $n=3$ and in the {\bf right panel} for $n=4$. We notice that for both cases, the post-inflationary EOS of the scalar field is stiff, that is $w_{\phi} \simeq 1$ for several e-folds after inflation.}
\label{fig:QI_exp_minus_l_EOS}
\end{center}
\end{figure}

\begin{figure}[t]
\begin{center}
\includegraphics[width=0.495\textwidth]{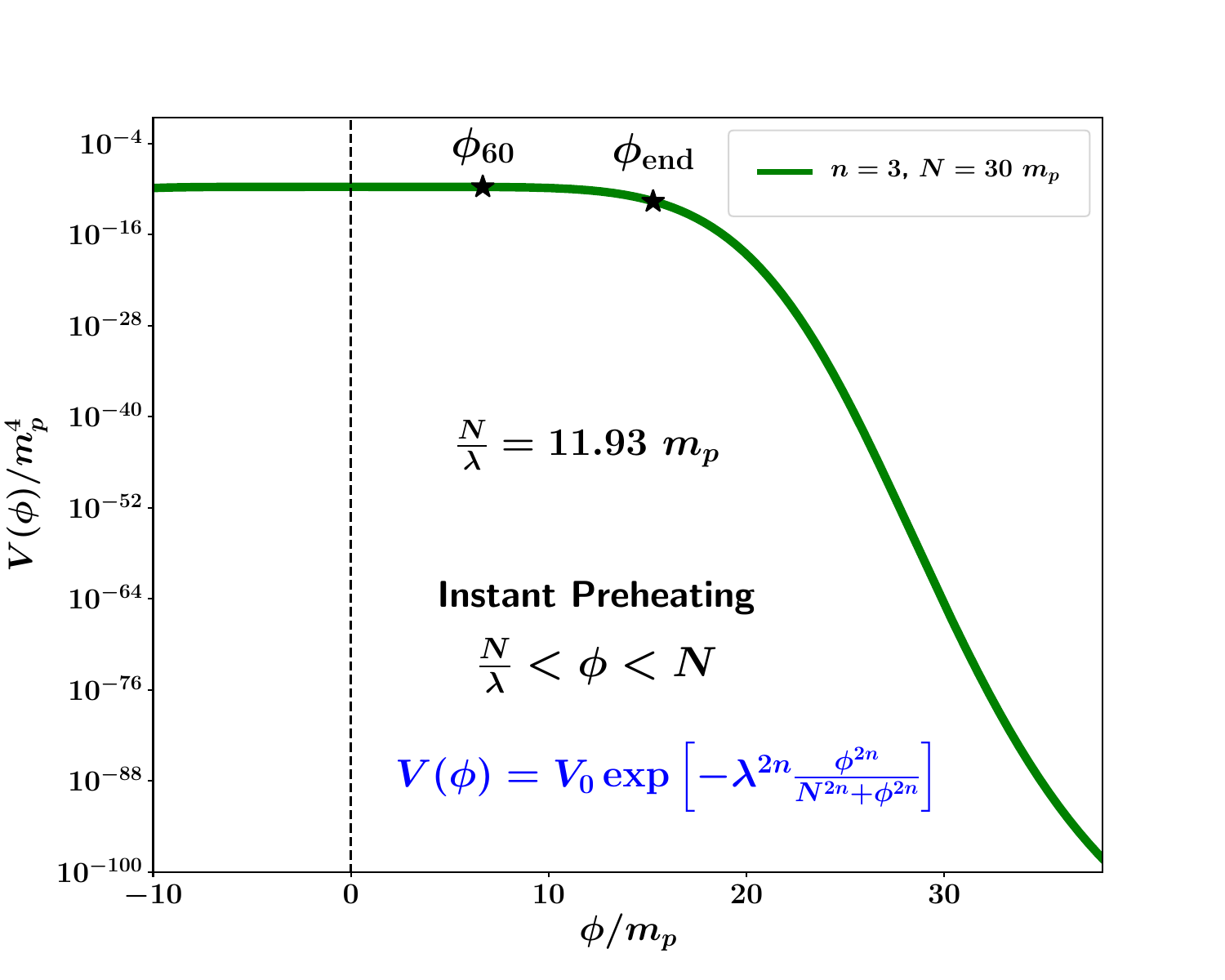}
\includegraphics[width=0.495\textwidth]{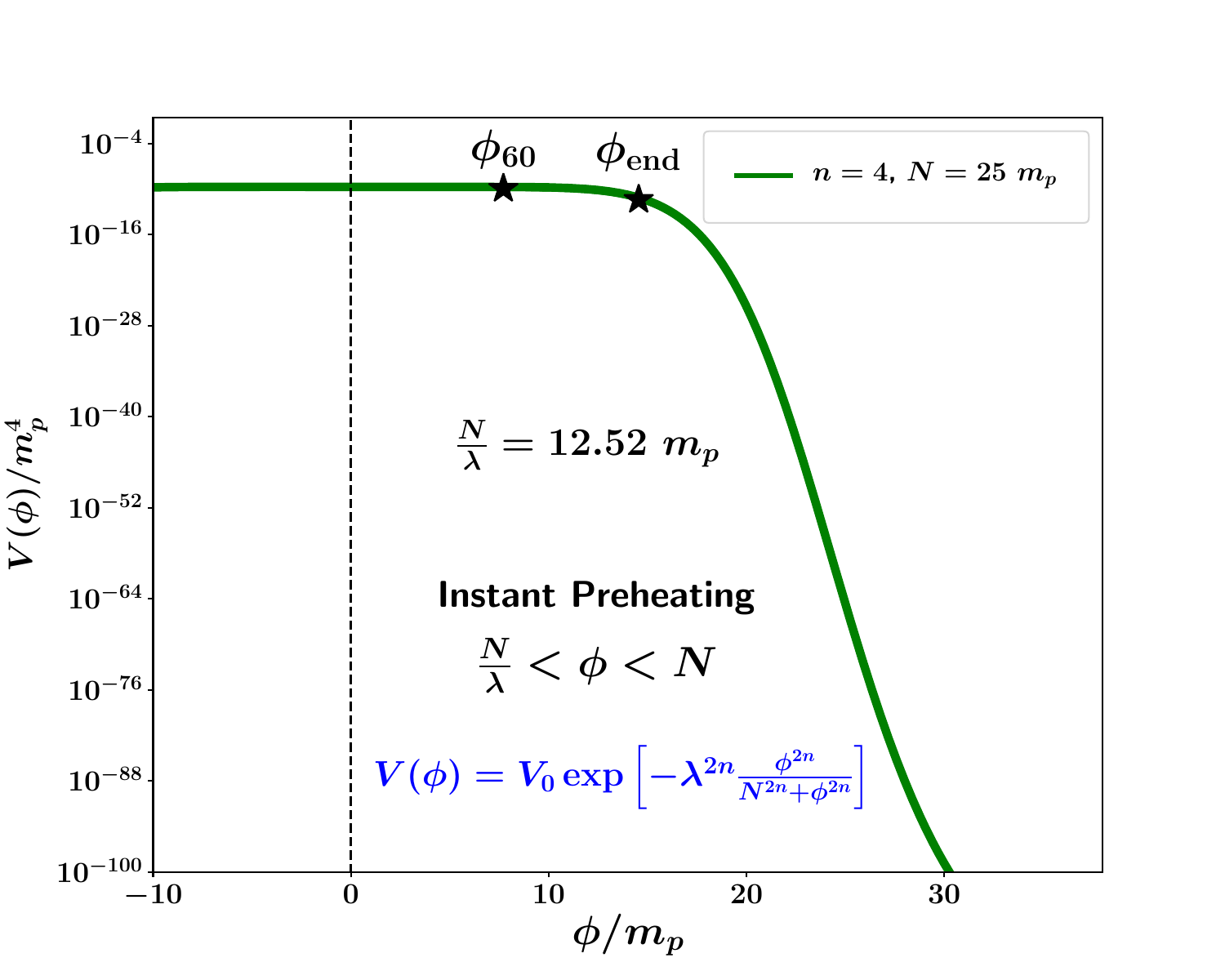}
\caption{This figure depicts the hilltop QI potential (\ref{eq:QI_exp_l_hilltop_genA}) in the {\bf left panel} for $n=3$ and in the {\bf right panel} $n=4$. The star marks indicate the field values corresponding to 60 e-folds (before the end of inflation) and the end of inflation. The potential steepens after the end of inflation and the inflaton field enters into a kinetic dominated phase  whereupon $w_\phi \simeq 1$ and
 the universe reheats via the mechanism of instant preheating.}
\label{fig:QI_exp_minus_l_pot}
\end{center}
\end{figure}

The EOS of the inflaton field is shown in Fig.~\ref{fig:QI_exp_minus_l_EOS} for $n=3,\, 4$. Fig.~\ref{fig:QI_exp_minus_l_pot} shows the potential (\ref{eq:QI_exp_l_hilltop_gen}) by illustrating  its  inflationary and post-inflationary regimes.

\bigskip

\begin{large}
\textbf{(B) Plateau QI:}
\end{large}
\medskip

\begin{figure}[htb]
\centering
\includegraphics[width=0.8\textwidth]{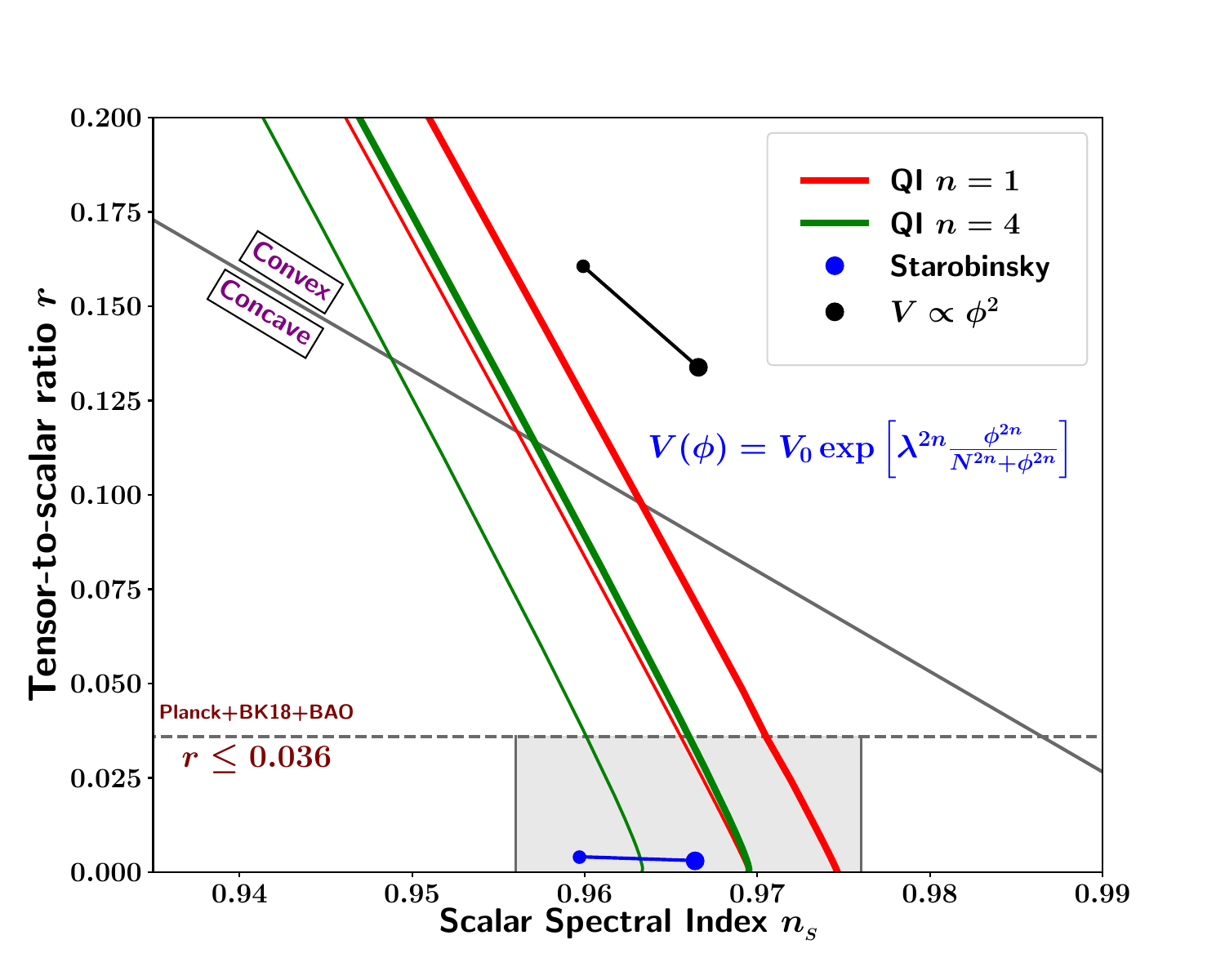}
\caption{The tensor-to-scalar ratio $r$ and scalar spectral tilt $n_{_S}$ are plotted for the plateau QI potential (\ref{eq:QI_exp_l_hilltop_genB}) for $n= 1,\,4,$ plotted in red and green colours respectively (thicker lines correspond to $N_e = 60$ while thinner lines correspond to $N_e = 50$, as per  the standard convention). The CMB 2$\sigma$ bound $0.957 \leq n_{_S} \leq 0.976$
 and the upper bound on  $r$, given in  Eq.~(\ref{eq:CMB_ns_r_constraint}), are
 indicated by the grey colour shaded region.}
\label{fig:QI_exp_plus_l_r_ns}
\end{figure}

As shown in section \ref{sec:QI_dual_exp}, the inverse potential of  (\ref{eq:QI_exp_l_hilltop_genA}), namely
\beq
V(\phi) = V_0 \, \exp\left[\frac{(\lambda\phi)^{2n}}{N^{2n} + \phi^{2n}}\right] \, ,
\label{eq:QI_exp_l_hilltop_genB}
\eeq
describes plateau QI.
Predictions for CMB observables $\lbrace n_{_S},r \rbrace$ are shown in Fig.~\ref{fig:QI_exp_plus_l_r_ns}. Note that this potential satisfies the CMB 2$\sigma$ constraints for $n= 1,\, 2, \, 3,\,4$. 
As briefly discussed in section 
\ref{sec:QI_dual_exp_lambda}, the inflaton field oscillates after the end of inflation 
allowing reheating to be sourced by the  decay of the coherently oscillating scalar field.

\section{Reheating in plateau and hilltop QI}
\label{sec:QI_reheating_dual}

An important aspect  of inflationary cosmology is the process of reheating. During reheating the inflaton field transfers its energy to other matter/radiative degrees of freedom whose subsequent decay, scattering and thermalization leads to the standard thermal hot Big Bang phase before the commencement of Big Bang Nucleosynthesis (BBN).  Below we provide a brief summary of  reheating in the context which is relevant for this work. The reader is referred to \cite{Kofman:1996mv,Kofman:1994rk,Kofman:1997yn,Shtanov:1994ce,Felder:1998vq,Felder:1999pv} for more details on the subject.

\subsection{Reheating in plateau QI potentials}
\label{sec:reheat_plateau}

The QI  potentials discussed in sections \ref{sec:QI_dual_KKLT}, \ref{sec:QI_dual_exp} and \ref{sec:QI_inf_dual},
 possess a minimum around which  the inflaton field oscillates  after the end of inflation 
(see Fig.~\ref{fig:pot_plateau_KKLT}). During these oscillations reheating can  occur either  perturbatively,
 due to the decay of individual massive inflaton particles, or non-perturbatively via the mechanism of   parametric resonance due to the coherent nature of the inflaton field oscillations. Which of these two ways is realized depends upon the nature of the coupling between the inflaton and other bosonic/fermionic degrees of freedom.
\begin{itemize}

\item {\bf Perturbative reheating}

The perturbative theory of reheating was originally developed in the context of the new inflationary
scenario in \cite{Albrecht:1982mp}.
Phenomenologically it amounts to adding a friction term $\Gamma {\dot\phi}$ to the classical
equation of  motion of the scalar field oscillating around the minimum of  its potential \cite{Albrecht:1982mp,Kolb:1990vq}
\beq
{\ddot\phi} + 3H{\dot\phi} + \Gamma {\dot\phi} + V'(\phi) = 0~.
\label{eq:EOM_damp}
\eeq
The perturbative theory of reheating works well if either (a) the inflaton decays only into fermions
$\psi$ through
a $h\psi{\bar\psi}\phi$ coupling with $h^2 \ll m_\phi/\mpl$, or (b)
the coupling of the inflaton to bosons, $\chi$, described by $\frac{1}{2}g^2\phi^2\chi^2$ is weak with
$g \ll 3\times 10^{-4}$, making particle production via parametric resonance ineffective \cite{Kofman:1996mv}.

In the perturbative regime, the reheating EOS is given by the average EOS of the inflaton field during its oscillations around the minimum of the potential \textit{i.e} $w_{\rm re} = \langle w_{\phi} \rangle$. For potentials which behave like $V(\phi) \propto \phi^{2n}$ close to their minimum value, the equation of state during reheating is given by
\beq
w_{\rm re} = \langle w_{\phi} \rangle = \f{n-1}{n+1}~.
\label{eq:w_reheat}
\eeq 
Reheating in the perturbative scenario is complete when the (decreasing) expansion rate
becomes equal to the decay rate so that $H \simeq \Gamma$. Following thermalization, the reheating temperature
is given by $T_{\rm re} \simeq 0.1\sqrt{\Gamma\mpl}$ which is independent of the duration of inflation and the properties of $V(\phi)$. The fact that the coupling between matter and the inflaton can alter, via radiative corrections, the shape of $V(\phi)$, places strong constraints on the total decay rate: $\Gamma < 10^{-20}\mpl$. This in turn implies that the reheating temperature  in perturbative models can be relatively small $T_{\rm re} < 10^9$ GeV and hence the duration of reheating $N_{\rm re}$ can be quite large \cite{Kofman:1996mv}.  As pointed out in \cite{Sahni:1990tx} and discussed  in  section \ref{sec:GW}, the spectrum of relic gravitational waves created during inflation is very sensitive to the post-inflationary
EOS,  $\langle w_{\phi} \rangle $. (As shown in \cite{Mishra:2021wkm} observations of the GW spectrum can 
help break degeneracies between different inflationary models.) 

\item {\bf Non-perturbative reheating}

For inflationary models in which the main source of reheating is through the decay of the inflaton
into bosons, the universe thermalizes and reheats through a sequence of successive stages.
\begin{enumerate}

\item
The first stage, sometimes called {\em preheating},
 sees the commencement of a parametric resonance brought about by coherent
oscillations of the inflaton $\phi$ around the minimum of its potential. The resonance can be
either narrow or broad depending upon (a) the value of coupling constant $g$ in the interaction
$\frac{1}{2}g^2\phi^2\chi^2$ between the inflaton and the bosonic field $\chi$, (b) the scalar field
amplitude $\Phi \equiv \phi_{\rm max}$, (c) its effective mass, $m_\phi^2 = V''$. 
If the resonance is broad ($g^2\Phi^2/m_\phi^2 \ggeq 1$)  then coherent oscillations of $\phi$
give rise to an exponentially large number of quanta of the field $\chi$ in a discrete
set of wave bands.
(The existence of self-interaction, such as the presence of
 a $\lambda \phi^4$ term in the Lagrangian, can also result in the creation
of quanta of the $\phi$ field during oscillations.)

\item
The second stage witnesses the backreaction of $\chi$ on $\phi$ (via scattering).
This effect can be quite significant
and can lead to the termination of the resonance. 

\item
During the third stage, which can be quite prolonged, quanta of the $\phi$ and $\chi$ fields
transfer their energy into other matter fields including radiation. The interaction between the
quanta of different fields leads to their thermalization and results in the universe acquiring
a reheating temperature $T_{\rm re}$.

\end{enumerate}
 Thus the end of the third stage leads to the commencement of
 the radiative hot Big Bang  phase of expansion during which the EOS in the universe is
$p \simeq \rho/3$.
The dynamics of the three stages of reheating is quite complex and usually requires a numerical
treatment \cite{Khlebnikov:1996wr}.
It is however quite instructive to characterize the pre-radiative stages
(1) - (3) by an effective EOS which, following \cite{Cook:2015vqa}, we assume to be a constant lying
in the interval $-1/3 < w_{\rm re} < 1$.

\end{itemize}

\subsection{Reheating in hilltop QI potentials}
\label{sec:reheat_hilltop}
  
    For hilltop QI models the potential does not possess a minimum but features a rapidly declining wing 
(see Fig.~\ref{fig:pot_hilltop_KKLT}). On this wing the inflaton field 
enters into a kinetic dominated fast rolling phase which leads to the
 post-inflationary EOS $w_{\phi} \simeq 1$ as shown in Fig.~\ref{fig:QI_exp_minus_l_EOS}.
Consequently reheating in this scenario can occur via either
 gravitational particle production or through the mechanism of instant preheating.
  
  The gravitational reheating mechanism  is slow and  inefficient, and  usually results in a  long duration of 
reheating $N_{\rm re}\geq 10$, with a low reheating temperature $T_{\rm re}$.
For our hilltop QI potentials which exhibit a stiff EOS after the end of  inflation,  the  long duration of
the kinetic regime may lead to an overproduction of post-inflationary gravitational waves  whose density violates the Big Bang  Nucleosynthesis bound $\left (\rho_{\rm GW}/\rho_r\right ) \lleq 0.2$ (as shown in \cite{Sahni:2001qp}).

Consequently this method of reheating is now seldom used in constructing models of QI
and most current models of QI usually rely on the 
mechanism of {\em instant preheating} \cite{Felder:1998vq,Felder:1999pv}.
In this scenario the inflaton field $\phi$ couples to a scalar field $\chi$ which in turn couples to a fermion $\psi$ with a farily large coupling constant.  The production of $\chi$ particles
commences when the effective mass of the scalar field $\chi$ begins to change
non-adiabatically after the end of inflation. Subsequently the field $\chi$ quickly decays into fermions and other standard model fields, thereby reheating the universe quite efficiently.

The instant preheating mechanism is therefore  frequently deployed in connection with models of hilltop
quintessential inflation (QI) in which the potential  usually steepens post-inflation and
results in the scalar field rolling rapidly with ${\dot\phi}^2 \gg V(\phi)$.
This leads to the equation of state $w_\phi \simeq 1$ and results in a blue spectrum for
gravitational waves, as originally shown in \cite{Sahni:2001qp}.
Preheating in the hilltop models introduced in this paper is further discussed in section \ref{sec:hilltop_heat}.

To summarize,
reheating in our models of quintessential inflation can occur in two distinct ways:
(i) For the plateau potentials (\ref{eq:QI_exp_plateau_gen}) \& (\ref{eq:QI_exp_l_plateau_gen}) reheating takes place
during oscillations of  $\phi$ around $\phi=0$, as discussed in \cite{Kofman:1994rk,Kofman:1997yn,Shtanov:1994ce}.
(ii) For the hilltop models described by Eqs.~(\ref{eq:QI_exp_hilltop1}) \& (\ref{eq:QI_exp_l_hilltop_gen})
the universe reheats via  instant preheating \cite{Felder:1998vq,Felder:1999pv}
 when $\phi$ rolls down the steep wing of its potential. 

As noted earlier, and discussed in detail in section \ref{sec:GW}, the pre-radiative EOS $w_{\rm re}$ 
during reheating 
sensitively affects the spectrum of relic gravitational waves produced in the early universe.
Future space-based GW experiments might therefore shed light on this
important parameter, and through it on the physics of the reheating epoch.

\section{Relic Gravitational Waves from QI}
\label{sec:GW}
Relic gravitational waves  are a
generic prediction of the inflationary scenario \cite{Starobinsky:1979ty}.
These  tensor fluctuations, which are created quantum mechanically,  get stretched to super-Hubble scales during inflation, where they remain frozen until their subsequent Hubble re-entry at late times after 
inflation ends. After becoming sub-Hubble, inflationary tensor fluctuations behave like a
stochastic relic gravitational waves background in the universe. The amplitude of relic GWs 
is sensitive to the value of the Hubble parameter
during inflation, while the GW spectrum encodes both the inflationary and post-inflationary
EOS of the universe \cite{Starobinsky:1979ty,Allen:1987bk,Sahni:1990tx,Sahni:2001qp,Souradeep:1992sm,Caprini:2018mtu,Figueroa:2019paj,Bernal:2019lpc}. Since GWs interact minimally with other forms of matter/radiation
they constitute one of the cleanest probes of the physics of the very early universe.

Transverse and traceless tensor fluctuations $h_{ij}$ during inflation are described by the action \cite{Maldacena:2002vr,Baumann:2018muz}
\beq
    S[h_{ij}] = \frac{1}{2} \int {\rm d}\tau~{\rm d}^3\vec{x}~\left(\frac{a m_p}{2}\right)^2 \left[ h_{ij}'^2 - (\partial h_{ij})^2\right] \, ,
\eeq
where $h_{ij}$ can be decomposed into its two orthogonal  polarization components,
\begin{align}
    h_{ij} &= \frac{1}{\sqrt{2}} \begin{bmatrix} h_+ & h_{\times} & 0 \\ h_{\times} & -h_+ & 0 \\ 0 & 0 & 0 \end{bmatrix}  = \frac{1}{\sqrt{2}} \begin{bmatrix} 1 & 0 & 0 \\ 0 & -1 & 0 \\ 0 & 0 & 0 \end{bmatrix} h_+ + \frac{1}{\sqrt{2}} \begin{bmatrix} 0 & 1 & 0 \\ 1 & 0 & 0 \\ 0 & 0 & 0 \end{bmatrix} h_{\times} \\
    &= \frac{1}{\sqrt{2}}~\epsilon^+_{ij} \,  h_+ + \frac{1}{\sqrt{2}}~\epsilon^{\times}_{ij} \,  h_{\times} \, ,
\end{align}
with $\epsilon^{\lambda}_{ij} \epsilon^{ij \, \lambda'} = 2 \delta^{\lambda \lambda'};~\lambda=\{+, \times \}$. The action can be rewritten as 
\beq
    S[h_+,\, h_\times] = \frac{1}{2} \int {\rm d}\tau~{\rm d}^3\vec{x}~\left(\frac{a m_p}{2}\right)^2 \, \sum_{\lambda=+,\times} \, \left[ ({h_\lambda}')^2 - (\partial h_\lambda)^2\right] \,
\eeq
which is equivalent to the action of two massless scalar fields in de Sitter spacetime.  We define  the two Mukhanov-Sasaki  variables for tensor type perturbations to be $v_\lambda = \left(a m_p/2 \right) h_\lambda$, in terms of which, the action becomes
\beq
    S[v_+,v_\times] = \frac{1}{2} \int {\rm d}\tau~{\rm d}^3\vec{x}~ \, \sum_{\lambda=+,\times} \, \left[ ({v_\lambda}')^2 - (\partial v_\lambda)^2 + \frac{a''}{a} {v_\lambda}^2\right] \, .
\eeq
The full power spectrum of tensor fluctuations is given by 
\begin{align}
    {\cal P}_T(k) &= \frac{k^3}{2 \pi^2} \left(\abs{h_+}^2 + \abs{h_{\times}}^2\right) = \frac{k^3}{2 \pi^2} \left(\frac{a m_p}{2}\right)^2 \left(\abs{v_+}^2 + \abs{v_{\times}}^2\right) \, ,
\end{align}
which (after imposing Bunch-Davies initial conditions) takes the following form on super-Hubble scales  (see Refs.~\cite{Baumann_TASI,Baumann:2018muz,Figueroa:2019paj})
\beq
 {\cal P}_T(k) =   \frac{2}{\pi^2} \left(\frac{H}{m_p}\right)^2 \, .
\label{eq:P_T}
\eeq
The tensor-to-scalar ratio is defined, in terms of ${\cal P}_T$, to be 
\beq
r = \f{{\cal P}_T}{{\cal P}_\zeta}~.
\label{eq:GW_r}
\eeq
The tensor power spectrum ${\cal P}_T(k)$ can be written  in the form of a power law as 
\beq
{\cal P}_T(k) = {\cal P}_T(k_*)\left (\frac{k}{k_*}\right )^{\nt}~,
\label{eq:GW_Ph1}
\eeq
where the tensor power  at the CMB pivot scale $k_* = 0.05 ~{\rm Mpc}^{-1}$ is  given, in terms of the scalar power,  by 
\beq
A_T \equiv {\cal P}_T(k_*) =  r \, A_{_S} =r\times 2.1 \times 10^{-9}~,
\label{eq:GW_Ah}
\eeq
and the tensor tilt is found to be 
\beq
 \nt = \frac{d\log{{\cal P}_T(k)}}{d\log{k}} = -\f{r}{8}~,
\label{eq:GW_nt}
\eeq
which satisfies the single-field consistency relation.

\subsection{Frequency of relic gravitational waves}
\label{sec:f_GW}

The  quantum mechanically generated tensor modes discussed above, which become super-Hubble  during inflation, make their Hubble re-entry at late times when $k=aH$ and behave like a stochastic GW
background \cite{Starobinsky:1979ty,Caprini:2018mtu,Figueroa:2019paj}.  The physical frequency of these stochastic GWs at the present epoch is given by

\beq
f = \f{1}{2\pi}\l(\f{k}{a_0}\r)=\f{1}{2\pi}\l(\f{a}{a_0}\r)H~,
\label{eq:GW_f}
\eeq 
where $a$, $H$ correspond to the scale factor and Hubble parameter of the universe during the epoch when the corresponding    tensor mode makes its Hubble re-entry. In this work, we focus on relic GWs that become sub-Hubble prior to 
matter-radiation equality, so that their characteristic frequency is large enough to enable them to be detected by the GW observatories in the near future. Expressing $H$ in terms of  temperature, we get 
\beq
\f{H}{\mpl} = \l(\f{\rho}{3\mpl^4}\r)^{\f{1}{2}} = \l(\f{\f{\pi^2}{30}\,g_T\,T^4}{3\mpl^4}\r)^{\f{1}{2}} = \pi \, \l(\f{g_T}{90}\r)^{\f{1}{2}} \, \l(\f{T}{\mpl}\r)^2~.
\label{eq:GW_H_T}
\eeq
Using entropy conservation (see \cite{Mishra:2021wkm}), one obtains
\beq
\f{a}{a_0} = \l(\f{a_{\rm eq}}{a_0}\r)\l(\f{g_{\rm eq}^s}{g_T^s}\r)^{1/3}\l(\f{T_{\rm eq}}{T}\r)~.
\label{eq:GW_a_T}
\eeq

Substituting $H$ from Eq.~(\ref{eq:GW_H_T}) and  $a/a_0$ from Eq.~(\ref{eq:GW_a_T}) in Eq.~(\ref{eq:GW_f}), we obtain the following important  expression for the present day frequency of GWs in terms of their Hubble re-entry 
temperature
\beq
f = 7.36\times 10^{-8}\, {\rm Hz}\,\l(\f{g_0^s}{g_T^s}\r)^{\f{1}{3}}\,\l(\f{g_T}{90}\r)^{\f{1}{2}}\,\l(\f{T}{{\rm GeV}}\r) ~.
\label{eq:GW_f_master}
\eeq
Values of $f$ corresponding to relic  GWs that became sub-Hubble at a number  of  important cosmic epochs are 
shown in table \ref{table:3}.

\begin{table}[htb]
\begin{center}
 \begin{tabular}{||c|c|c|c|c||} 
 \hline\Tstrut
 \bf{Epoch} & \bf{Temperature} $T$ & \bf{GW Present day  {\large f} (in Hz)}\\ [1ex] 
 \hline\hline\Tstrut
 
 Matter-radiation equality & $\sim 0.8$~eV  & $1.14\times 10^{-17}$ \\ [1.2ex] 
 \hline\Tstrut
 CMB pivot scale re-entry & $\sim 5$~eV  & $8.5\times 10^{-17}$ \\ [1.2ex]
 \hline\Tstrut
 Big Bang Nucleosynthesis & $\sim 1$~MeV  & $1.8\times 10^{-11}$ \\ [1.2ex]
  \hline\Tstrut
 QCD phase transition & $\sim 150$~MeV  & $2.95\times 10^{-9}$ \\ [1.2ex]
 \hline\Tstrut
 Electro-weak symmetry breaking & $\sim 100$~GeV  & $2.7\times 10^{-6}$ \\ [1.2ex]
 \hline
\end{tabular}
\captionsetup{
	justification=raggedright,
	singlelinecheck=false
}
\caption{Present day frequencies of relic GWs have been tabulated  for five different temperature  scales, associated with the Hubble re-entry of the respective primordial tensor modes. In order to probe
 the epoch of reheating using relic GWs, the physical frequency corresponding to tensor modes which 
become sub-Hubble during reheating must satisfy  $f>f_{_{\rm BBN}} \simeq 10^{-11}$ Hz 
so that reheating terminates before the commencement of BBN. }
\label{table:3}
\end{center}
\end{table}

\subsection{Spectral energy density of relic gravitational waves}
\label{sec:Omega_GW}

The present day spectral density of stochastic GWs, defined in terms of the  critical density at the present epoch $\rho_{_{0c}}$  by \cite{Starobinsky:1979ty,Caprini:2018mtu,Figueroa:2019paj}

\beq
\Og (f) \equiv \f{1}{\rho_{_{0c}}}\f{d\rho_{\rm GW}^0(f)}{d\log{f}}~,
\label{GW_Omega_def}
\eeq
is given  by the following set of equations \cite{Sahni:1990tx,Sahni:2001qp,Figueroa:2019paj,Ema:2020ggo,Mishra:2021wkm}
\ber
\Og^{(\rm RD)}(f) &=& \frac{1}{24} \, {\cal P}_T(f)\,\Omega_{0r},
~ f_{\rm eq} < f \leq f_{\rm re}\label{eq:GW_spectrum_1b}\\
\Og^{(\rm re)} (f) &=& \Og^{(\rm RD)}
\left (\frac{f}{f_{\rm re}}\right )^{2\left (\frac{w-1/3}{w+1/3}\right )},
~ f_{\rm re} < f \leq f_{\rm e}
\label{eq:GW_spectrum_1c}
\eer
where $f_{\rm eq}$, $f_{\rm re}$, $f_e$ refer  to the present day frequency of relic GWs corresponding to tensor modes that became sub-Hubble at:  
the epoch of matter-radiation equality ($f_{\rm eq}$), at the end of reheating 
(commencement  of the radiation dominated epoch, $f_{\rm re}$) and at the end of inflation ($f_e$). 
The superscripts `RD' and  `re' in $\Og$ refer to the radiative epoch and the epoch of reheating respectively. For $\Omega_{0r}=1$, Eq.~(\ref{eq:GW_spectrum_1b}) coincides with the results of \cite{Starobinsky:1979ty}. 
Note that $f_{\rm re}>f_{\rm BBN} \simeq 10^{-11}~{\rm Hz}$ in order to 
satisfy the BBN bound on $\Tre$. As concerns the EOS $w = \wre$ during the epoch of reheating, it is important to keep in mind the following points:

\begin{itemize}

\item
In the case of perturbative reheating  for $V(\phi)\propto \phi^{2n}$, the value of $\wre \equiv \l<w_{_{\phi}}\r>$
is given by $\l<w_{_{\phi}}\r> = \frac{n-1}{n+1}$.

\item In the case of non-perturbative reheating, the physics of the reheating
epoch can be quite complex. In this case $\wre$ is sometimes assumed 
to be a constant, for the sake of simplicity \cite{Cook:2015vqa,Dai:2014jja,Munoz:2014eqa}.

\end{itemize}

From the equations developed earlier in this section and from Eqs.~(\ref{eq:GW_spectrum_1b}),(\ref{eq:GW_spectrum_1c}),
  it follows that the 
spectral density of stochastic GWs corresponding to modes that became sub-Hubble prior to 
matter-radiation equality is
\ber
\mbox{{\bf Radiative epoch:}} ~~   \Og^{(\rm RD)}(f)  = \l(\frac{1}{24}\r)\,r\,A_{_S}\,\l(\f{f}{f_*}\r)^{\nt}\,\Omega_{0r} ~,  ~~ f_{\rm eq} < f \leq f_{\rm re}\, ,
\label{eq:GW_spectrum_2a}\\
\mbox{{\bf During reheating:}} ~~ \Og^{(\rm re)} (f)  = \Og^{(\rm RD)}(f)\, \left (\frac{f}{f_{\rm re}}\right )^{2\left (\frac{w-1/3}{w+1/3}\right )},~~ f_{\rm re} < f \leq f_{\rm e}\, ~,
\label{eq:GW_spectrum_2b}
\eer
where we have used ${\cal P}_T(f) = {\cal P}_T(f_*)\left (\frac{f}{f_*}\right )^{\nt}$, with $A_T \equiv {\cal P}_T(f_*) =  r \, A_{_S}$ from Eq.~(\ref{eq:GW_Ah}). Note that 
$f_*$ is the physical frequency (of GW) corresponding to the CMB pivot scale comoving wave number $k_*$.  

Eqs~(\ref{eq:GW_Ph1}), (\ref{eq:GW_spectrum_1b}), (\ref{eq:GW_spectrum_1c}) allow us to define a local post-inflationary
gravitational wave (tensor) spectral index as follows
\beq
\ng = \frac{d\log{\Og(k)}}{d\log{k}} = \frac{d\log{\Og(f)}}{d\log{f}}
\eeq
where
\beq
\ng =\nt + 2\,\left (\frac{w-1/3}{w+1/3}\right )
\label{eq:index1}
\eeq
which implies $\ng > \nt$ for $w > 1/3$, $\ng = \nt$ for $w = 1/3$ and
$\ng < \nt$ for $w < 1/3$,
where $w$ is the background EOS and is given by $w=0$ during matter domination,
$w=1/3$ during radiation domination and by $w = w_{\rm re}$ during reheating.

\begin{figure}[t]
\begin{center}
\includegraphics[width=0.68\textwidth]{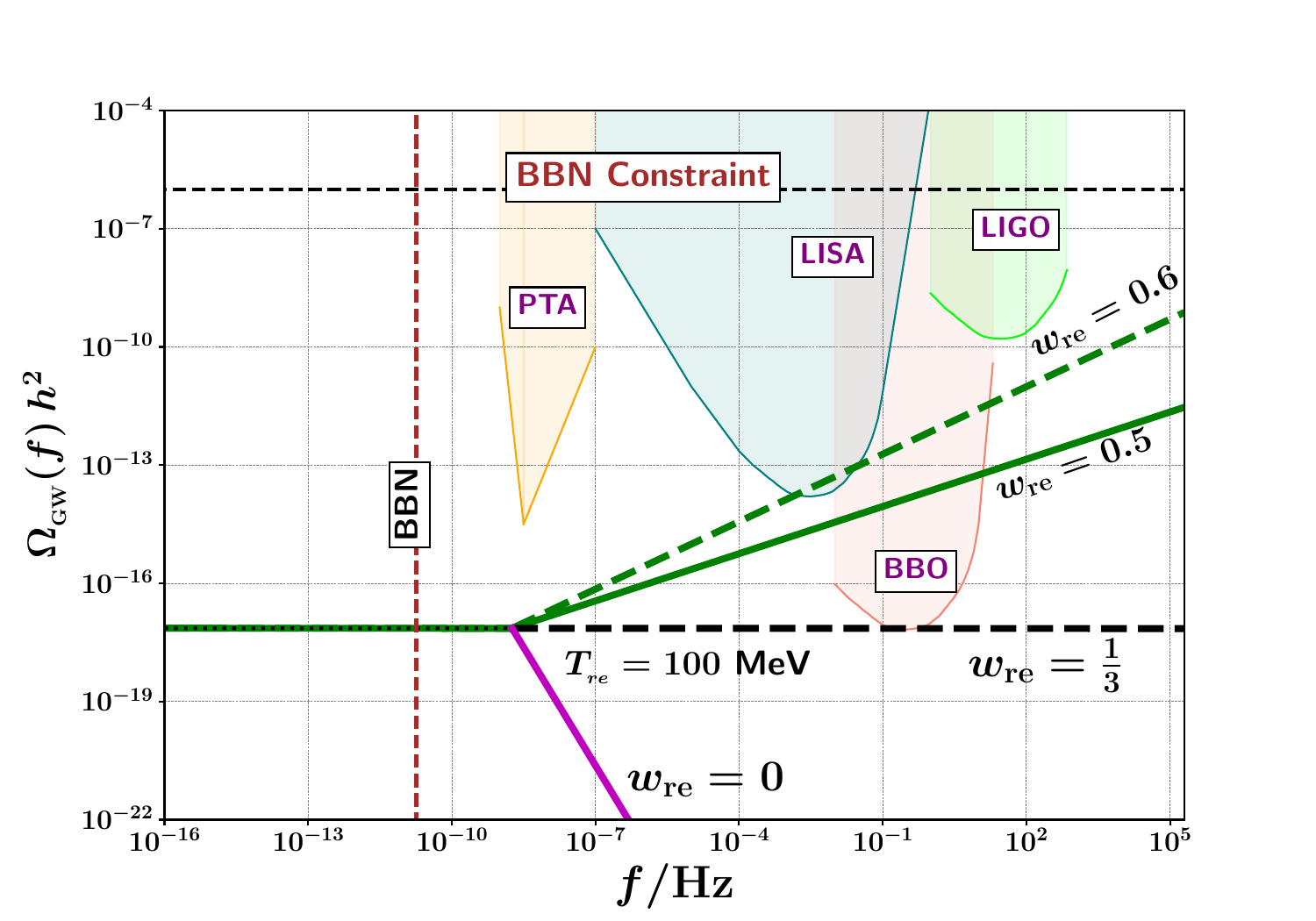}
\includegraphics[width=0.68\textwidth]{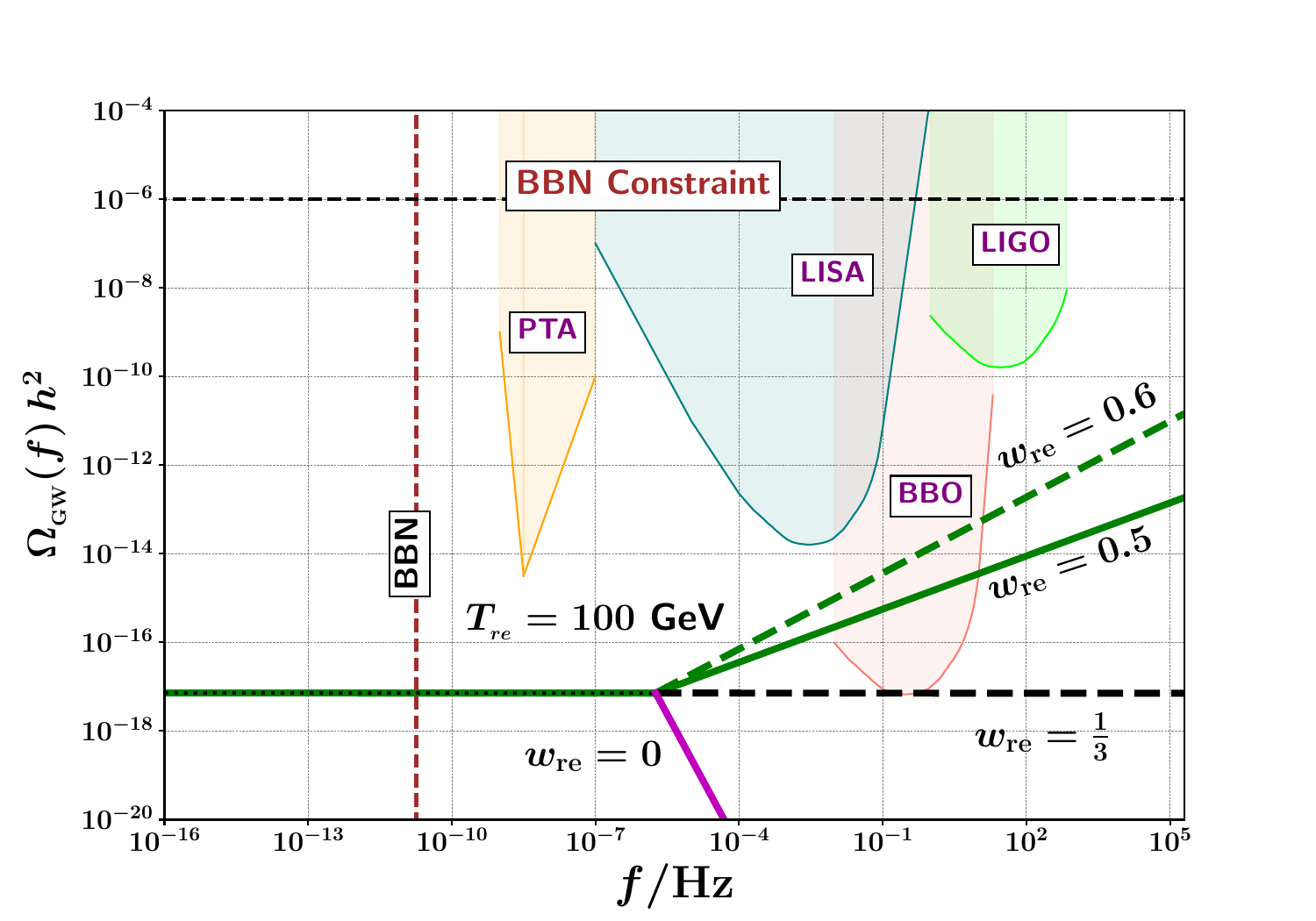}
\caption{The spectral energy density of relic gravitational waves is shown for plateau QI models  from  the perspective of 
 ongoing and near future GW observatories such as the advanced LIGO, LISA and BBO. {\bf Top panel} depicts the spectrum of  relic GWs corresponding  to reheating EOS $\wre = 0, ~1/3, ~1/2, ~3/5$, plotted in solid purple, dashed black, solid green  and dashed  green curves respectively, for  a fixed reheating temperature $\Tre=100$ MeV and tensor-to-scalar ratio $r\simeq 0.003$. 
The {\bf bottom panel} shows the same but with the higher reheating temperature  $\Tre=100$ GeV. }
\label{fig:GW_plateau}
\end{center}
\end{figure}

Note that since $\nt \simeq -2\epsilon_H$, 
CMB constraints on  the tensor-to-scalar ratio $r = 16\,\epsilon_H \leq 0.036$, imply 
$\vert \nt \vert \leq 0.0045$ (see Ref.~\cite{Mishra:2022ijb}). Hence $\nt$  is a very small quantity that does not generate an appreciable change in $\Og(k)$ over a 30 order of magnitude variation in $k$ (and hence in $f$). Therefore Eq.~(\ref{eq:index1}) effectively reduces to
\beq
\ng \simeq 2\,\left (\frac{w-1/3}{w+1/3}\right )~.
\label{eq:spectral_index}
\eeq
Thus the post-inflationary EOS has a direct bearing on the spectral index of relic gravitational
radiation with 
\ber
\ng \geq 0 ~~ {\rm for} ~~ w > 1/3\nonumber\\
 \ng \simeq 0 ~~ {\rm for} ~~ w = 1/3 \nonumber\\
\ng \lleq 0 ~~ {\rm for} ~~ w< 1/3
\label{eq:GW_ng_cases}
\eer
 which illustrates the extreme sensitivity of the 
GW spectral index to the background EOS in the universe.

\begin{figure}[t]
\begin{center}
\includegraphics[width=0.85\textwidth]{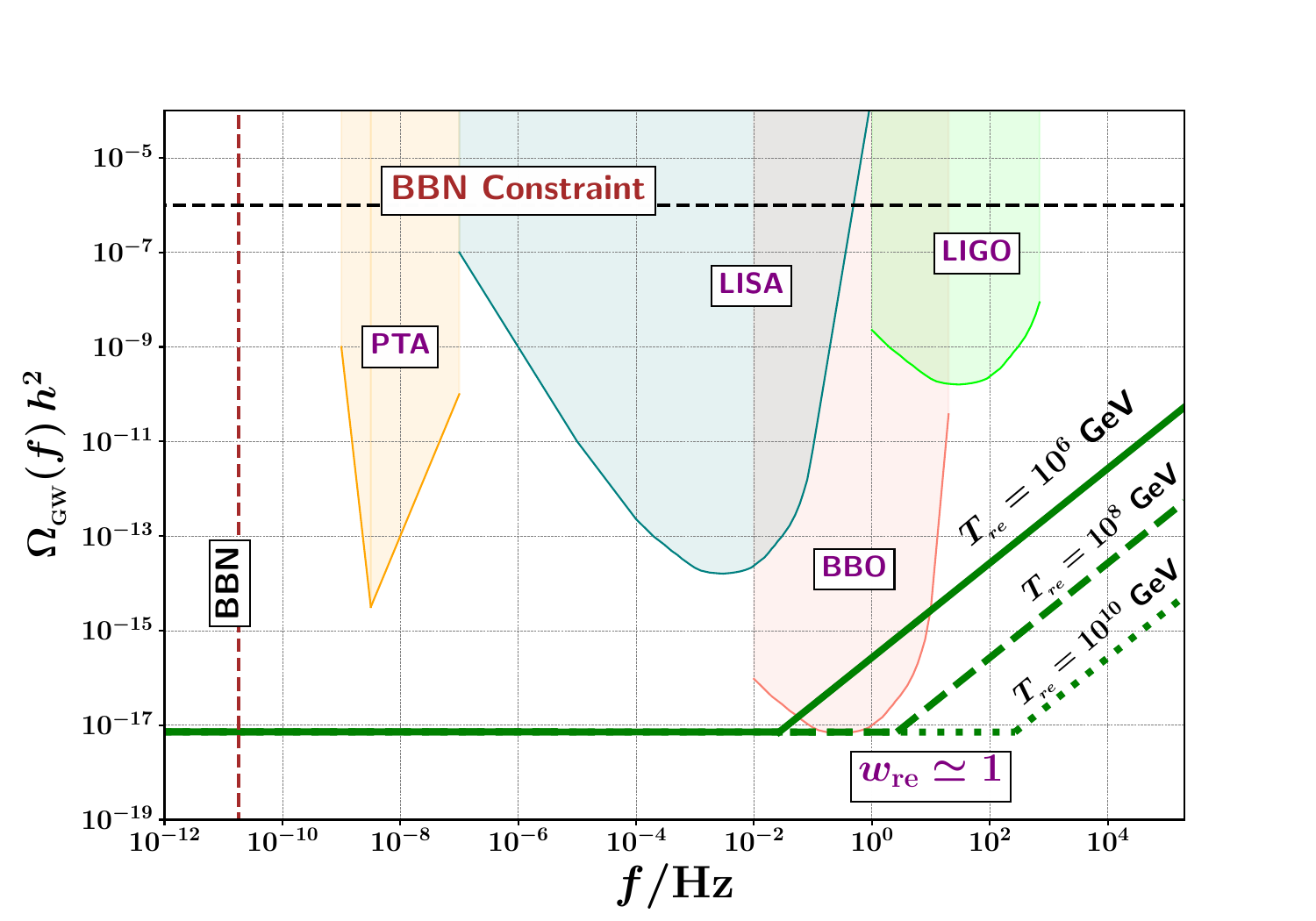}
\caption{
The spectral energy density of relic gravitational waves is shown for hilltop QI models 
in which the post-inflationary EOS during reheating is $\wre\simeq 1$ and the
tensor-to-scalar ratio is $r\simeq 0.003$. 
The solid, dashed  and dotted green curves correspond to reheating
temperatures $\Tre=10^6$ GeV, $\Tre=10^8$ GeV, and $\Tre=10^{10}$ GeV respectively. Note that for the hilltop models, a reheating temperature $\Tre < 10^5$ GeV will violate the  BBN constraint $\Og\, h^2 \leq 10^{-6}$. Theories predicting such signals can therefore be regarded as being unphysical.}
\label{fig:GW_hilltop}
\end{center}
\end{figure}

Setting $\nt = 0$ for simplicity, one gets, for the different cosmological epochs, the result
\ber
{\bullet} ~~~~ {\rm Matter ~ domination}~(w=0) ~~~~ \Rightarrow ~~ \ng(k)\bigg\vert_{\rm MD} &=& ~ - 2 \nonumber\\
{\bullet} ~~~~ {\rm Radiation ~ domination}~(w=1/3) ~~~~ \Rightarrow ~~ \ng(k)\bigg\vert_{\rm RD} &\simeq& ~~0
\eer

\n
$\bullet$ ~ During the pre-radiation epoch the GW spectrum depends upon the EOS during reheating.
In the context of perturbative reheating, during coherent oscillations of the inflaton around the potential $V(\phi)\propto \phi^{2n}$, one finds
\beq
\ng(k)\bigg\vert_{\rm OSC} = 2\l(\frac{n-2}{2n-1}\r)
\eeq

In marked contrast to perturbative reheating, in models with non-perturbative reheating the 
reheating/preheating epoch can be a complex affair with explosive (resonant) particle production, backreaction
and non-equilibrium field theory all playing a significant role until thermalization is finally reached. For
the sake of  simplicity this epoch is usually characterised (see \cite{Dai:2014jja,Creminelli:2014fca,Munoz:2014eqa,Cook:2015vqa}) by a constant effective EOS parameter, $\wre$, so that the general
formulae (\ref{eq:GW_spectrum_2a}) --  (\ref{eq:index1})    
also have bearing on this scenario.

The  spectral energy density of  GWs in  our plateau models of QI  has been illustrated  in Fig.~\ref{fig:GW_plateau} in the light of both ongoing and  near future GW observatories. The top panel shows the spectrum of  relic GWs corresponding  to reheating EOS $\wre = 0, ~1/3, ~1/2, ~3/5$, plotted in solid purple, dashed  black, solid green and dashed green curves respectively, for  a fixed reheating temperature $\Tre=100$ MeV and tensor-to-scalar ratio $r\simeq 0.003$. The bottom panel depicts the same but with a higher reheating temperature  $\Tre=100$ GeV.

The  spectral energy density of  GWs  for  hilltop QI models with $\wre \simeq 1$  is shown
  in Fig.~\ref{fig:GW_hilltop} for a fixed tensor-to-scalar ratio $r\simeq 0.003$. The solid, dashed and dotted green curves correspond to  reheating temperatures $\Tre=10^6$ GeV, $\Tre=10^8$ GeV, and $\Tre=10^{10}$ GeV respectively.

\section{Discussion}
\label{sec:discussion}

Owing to the  remarkable fact that the universe appears to accelerate twice: once at the early times during cosmic inflation and later at the late times during dark energy domination, numerous attempts have been made in the literature to describe both the phases of accelerated expansion of space under a common  framework. Such an unification  scheme, known as `Quintessential Inflation' (QI), is usually developed in the framework of a single  scalar field  $\phi$ with a suitable potential $V(\phi)$.  In order to facilitate both inflation and the late time acceleration, the  potential $V(\phi)$ must possess sufficiently flat wings both at early and late times. However the quintessential inflationary framework often incorporates unnaturally small parameters in the potential to obtain a low energy density $V_{\rm DE}$ of dark energy   from the high energy inflationary scale $V_{\rm inf}$, given the enormous difference between the two energy scales. In this work we have introduced new models of QI that successfully unify inflation and DE without incorporating unnaturally small parameters in the potential. 

         The QI models introduced in this work feature  plateau and flat-winged hilltop potentials that are related to each other through an inverse transformation. If  $V(\phi) = V_0 \, v(\phi)$   is a plateau QI potential, then the inverse $V(\phi) = V_0 \l[v(\phi)\r]^{-1}$ describes a flat-winged hilltop potential. In section \ref{sec:QI_dual_KKLT} we introduced KKLT-inspired  plateau and hilltop QI models based on the function $v(\phi) = \left\lbrack \frac{M^{2n} + \phi^{2n}}{N^{2n} + \phi^{2n}}\right\rbrack$ (and its inverse $\l[v(\phi)\r]^{-1}$) with $n \geq 1$.  In order to successfully source QI, the parameters $M$ and $N$ satisfy $M\ll N$ and are related by $\l(\f{M}{N}\r)^{2n} = \f{V_{\rm DE}}{V_{\rm inf}}$. Inflationary dynamics of this model was discussed in section \ref{sec:QI_inf_dual_KKLT}. The plateau QI potential (\ref{eq:plateau_KKLT}) satisfies the CMB constraints given in Eq.~(\ref{eq:CMB_ns_r_constraint}) on $\lbrace n_{_S},r \rbrace$ for $1 \leq n \leq 4$ as shown in Fig.~\ref{fig:KKLT_r_ns}. The inflaton field oscillates after the end of inflation and the universe reheats due to the decay of the oscillating inflaton field. At late times the  potential approaches a constant value $V(\phi) \rightarrow V_{\rm DE}$. Thus the plateau QI potential (\ref{eq:plateau_KKLT}) represents a viable model of QI. 
         
         However the hilltop QI potential (\ref{eq:hilltop_KKLT}) discussed in section \ref{sec:QI_hilltop_KKLT} does not satisfy the CMB constraints of Planck 2018 due to its small scalar spectral index $n_{_S} < 0.948$ (and it leads to extremely low  tensor-to-scalar ratio $r \ll 10^{-3}$). The potential also suffers from a fine tuning of initial conditions for inflation which plagues all small-field hilltop models. Therefore the hilltop  potential (\ref{eq:hilltop_KKLT}) does not lead to an appealing model of quintessential inflation.
         
          In section \ref{sec:QI_dual_exp} we introduced exponential based  plateau and hilltop QI models of the functional form $V(\phi) = V_0 \, e^{\pm f(\phi)}$, with $f(\phi)$ being an inflationary plateau or hilltop function. We discussed two concrete examples of this class of exponential based QI potentials. The first one is based on $f(\phi) = \exp{\left\lbrack \mp \frac{M^{2n}}{N^{2n} + \phi^{2n}}\right\rbrack}$ and the second one is based on  $f(\phi) = \exp{\left\lbrack \pm \frac{\l(\lambda \phi\r)^{2n}}{N^{2n} + \phi^{2n}}\right\rbrack}$. Both the models are equivalent to each other upon a parameter redefinition  $\f{M}{N} \longrightarrow \lambda$ as shown in appendix \ref{sec:app_eqv}. Hence we discuss the inflationary dynamics and CMB constraints on one of these  models, namely $f(\phi) = \exp{\left\lbrack \pm \frac{\l(\lambda \phi\r)^{2n}}{N^{2n} + \phi^{2n}}\right\rbrack}$. The exponential based hilltop QI potential (\ref{eq:QI_exp_l_hilltop_gen}) satisfies CMB constraints on $\lbrace n_{_S},r \rbrace$ for $n = 3,\, 4,$ as discussed in section \ref{sec:QI_inf_dual_exp_KKLT} and illustrated in Fig.~\ref{fig:QI_exp_minus_l_r_ns}. The exponential based plateau QI potential (\ref{eq:QI_exp_l_plateau_gen}) satisfies CMB constraints for $n =1, \, 2, \,  3,\, 4,$ as shown in Fig.~\ref{fig:QI_exp_plus_l_r_ns}. Hence both the plateau and the  hilltop potentials in this case represent viable models of QI.
         
        In section \ref{sec:QI_reheating_dual}, we discussed  the possible reheating mechanisms in our plateau and hilltop models of QI. Plateau models featuring potentials with a minimum leads to  coherent oscillations of the inflaton field after inflation. During this oscillatory phase, the inflaton decays  either perturbatively or non-perturbatively via parametric resonance \cite{Kofman:1996mv,Kofman:1994rk,Kofman:1997yn,Shtanov:1994ce} and the universe reheats successfully. In the hilltop QI models, the inflaton potential declines rapidly leading to a stiff  post-inflationary EOS \textit{i.e} $w_{\phi} \simeq 1$ as shown  in Fig.~\ref{fig:QI_exp_minus_l_EOS}.  In this case the universe is expected to reheat via the mechanism of instant preheating \cite{Felder:1998vq,Felder:1999pv}. Forecast for the gravitational wave background generated in these QI models was discussed in section \ref{sec:GW}. The spectral energy density of relic gravitational waves (\ref{GW_Omega_def}) is sensitive  to the post-inflationary EOS of the universe during reheating as shown in Eq.~(\ref{eq:GW_spectrum_2b}). Since the post-inflationary EOS is stiff-like for hilltop QI models, the associated GWs exhibit a blue tilt at higher frequencies as illustrated in Fig.~\ref{fig:GW_hilltop}. While for plateau QI potentials,  the spectral tilt of the  relic GWs can be red, flat or blue depending upon the effective reheating EOS during the scalar field oscillations and decay.
        
        We further introduced a number of possible quintessential inflationary potentials, including the non-flat QI potentials in section \ref{sec:QI_dual_nonflat}, a logarithm based QI potential in section \ref{sec:QI_log} and other additional QI potentials in appendix \ref{sec:appendix}.

\section{Acknowledgements}

We are thankful to the late Alexei A. Starobinsky for collaboration and comments during the earliest stage of this project and we dedicate this paper in his loving memory. 
SSM is supported by  STFC Consolidated Grants [ST/T000732/1 and ST/X000672/1]. SSM is also thankful to IUCAA, Pune for the hospitality.  
VS was partially supported by the J.~C.~Bose Fellowship of Department of Science and Technology, Government of India. For the purpose of open access, the
authors have applied a CC BY public copyright licence to any Author Accepted Manuscript version arising.

\medskip

{\bf Data Availability Statement:} This work is entirely theoretical and has no associated data.

\section*{Appendices}
\appendix

\section{Preheating in hilltop QI}
\label{sec:hilltop_heat}

As discussed in section \ref{sec:reheat_hilltop}, the mechanism of {\em instant preheating} is  frequently deployed in connection with models of hilltop
quintessential inflation (QI) in which the potential  usually steepens post-inflation and
results in the scalar field rolling rapidly with ${\dot\phi}^2 \gg V(\phi)$.
This leads to the equation of state $w_\phi \simeq 1$ and results in a blue spectrum for
gravitational waves.

\subsection{Inflaton decay for hilltop QI potentials}
\label{sec:IP_QI}

\begin{figure}[htb]
\begin{center}
\includegraphics[width=0.8\textwidth]{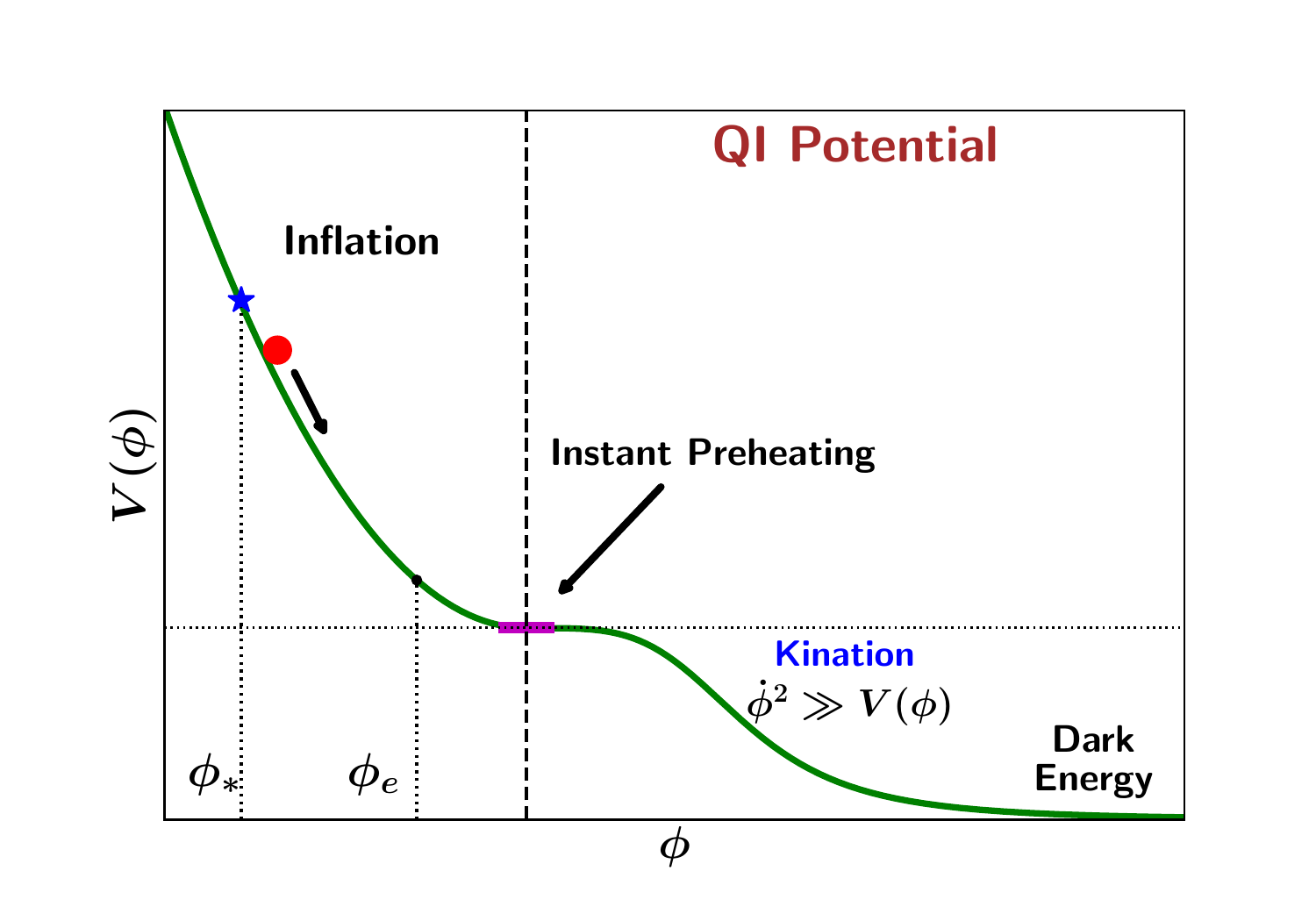}
\caption{This figure schematically illustrates a typical quintessential inflationary potential for which the mechanism of instant preheating works very well. Inflation happens in the left wing for $\phi < \phi_e <0$. Non-adiabatic particle production happens for a small range of field values  around $\phi=0$ (as indicated by the magenta colour region). After the end  inflation, the  potential rapidly declines for $\phi>0$ leading to a kination regime. For large positive values of $\phi$, the potential behaves like dark energy.}
\label{fig:pot_QI_PV}
\end{center}
\end{figure}

The theory of instant preheating was introduced in  \cite{Felder:1998vq,Felder:1999pv} for quintessential inflationary potentials satisfying the following properties (see Fig.~\ref{fig:pot_QI_PV})

\begin{enumerate}
\item An inflationary wing for $\phi < 0$. In the early works \cite{Peebles:1998qn}, the inflationary wing was assumed to be of the form $V(\phi) \simeq \f{1}{2}m^2\phi^2$ or $\lambda \phi^4$ (for $\phi<0$).
\item Inflation ends at some point $\phi=\phi_e < 0$.

\item After the end of inflation, the inflaton field rolls towards the right and crosses the point $\phi = 0$.

\item The potential possesses a rapidly declining wing for $\phi \geq 0$ leading to a phase of kinetic regime where $\dot{\phi}^2 \gg V(\phi)$ and hence $w_{\phi} \simeq 1$. 
\item At late times for large field values $\phi > 0$, the potential flattens and behaves like dark energy.   
\end{enumerate}

In this class of  models, the inflaton field $\phi$ couples to a scalar field $\chi$ (which in turn couples to a fermion $\psi$) through the interaction
Lagrangian 
\beq
{\cal L}_{\rm int} = - \frac{1}{2}g^2\phi^2\chi^2 - h \bar{\psi} \psi \chi
\label{eq:QI_L_int}
\eeq
and the production of $\chi$ particles
commences when the effective mass of the scalar field $m_\chi = g|\phi|$ begins to change
non-adiabatically so that
\beq
|{\dot m_\chi}| \geq m_\chi^2 ~~\Rightarrow ~~ |{\dot \phi}| \geq g\phi^2~.
\label{eq:QI_adiabaticity}
\eeq
 It is easy to see that violation of the  adiabaticity condition  naturally happens when the inflaton field crosses $\phi = 0$ where a small fraction $`f$' of the inflaton energy density gets converted to $\chi$ particles. The field $\chi$ then quickly decays into fermions $\psi$ owing to  a relatively large coupling $h \simeq 1$.  The fermions behave relativistically and their subsequent scattering leads to thermalization. Since the inflaton field  is kinetic dominated, it behaves like stiff matter with $w_{\phi} \simeq 1$. Consequently the inflaton density evolves as $\rho_{\phi} \propto 1/a^6$ while the density of the relativistic  decay products evolves as $\rho_r \propto 1/a^4$. Eventually the universe becomes radiation dominated and reheating is completed.
 
\subsection{Duration of Reheating}
\label{sec:Nre}
  The mechanism of instant preheating leads to an efficient way of reheating of the universe for QI potentials through the chain reaction $\phi \longrightarrow \chi \longrightarrow \psi$. By assuming that the $\chi$ particles immediately decay into fermions $\psi$, the duration of reheating can be computed in the following way.
  
             Suppose $a_{_{\rm IP}}$, $a_{\rm re}$ correspond to the  instant of creation of $\chi$ particles and the epoch of  reheating respectively. Let us assume that at $a = a_{_{\rm IP}}$, a fraction $f$ of the inflaton energy density gets converted into $\chi$ particles (which immediately decay into $\psi$ particles). Hence the density of the relativistic decay products at the instant of creation is given by   
             
\beq
\rho_r\bigg\vert_{\rm IP} =  f \,  \times  \rho_{\phi}\bigg\vert_{\rm IP}~.
\label{eq:frac}
\eeq  
 Reheating is completed when            
 
\beq
\rho_r\bigg\vert_{\rm re} =  \rho_{\phi}\bigg\vert_{\rm re}~,
\label{eq:rec}
\eeq  

$$\Rightarrow \rho_r\bigg\vert_{\rm IP} \l(\f{a_{_{\rm IP}}}{a_{\rm re}} \r)^4 = \rho_{\phi}\bigg\vert_{\rm IP} \l(\f{a_{_{\rm IP}}}{a_{\rm re}} \r)^6~,$$
leading to
\beq
\f{a_{\rm re}}{a_{_{\rm IP}}} = f^{-\f{1}{2}}~.
\label{eq:a_frac}
\eeq  

Hence the duration of reheating is given by

\beq
\Nre = \log_e{\l(\f{a_{\rm re}}{a_{_{\rm IP}}}\r)} = \f{1}{2}\log_e{\l(\f{1}{f}\r)}~.
\label{eq:Nre}
\eeq  

Usually the fraction of energy conversion from the inflaton $\phi$ to $\chi$ is given by $f\simeq 10^{-2} \, g^2$ (see \cite{Felder:1998vq}). Incorporating this into Eq.~(\ref{eq:Nre}), we obtain

\beq
\Nre = \log_e{\l(\f{10}{g}\r)}~.
\label{eq:Nre_1}
\eeq  

Since efficient reheating requires $g > 10^{-4}$, we obtain the following upper bound on the duration of reheating

\beq
\Nre \leq 5\, \log_e{(10)} = 11.5~.
\label{eq:Nre_bound}
\eeq 

An upper bound on the duration of reheating results in a lower bound on reheating temperature which can avoid the problem of over production of relic gravitational waves in the scenario of instant preheating.

\subsection{Instant preheating for our hilltop QI models}
\label{sec:IP_QI_hilltop}

While mechanism for instant preheating works very well for the class of QI models (with potential illustrated in Fig.~\ref{fig:pot_QI_PV}) where the inflaton crosses $\phi = 0$ post-inflation, it might not be realised in QI models where $\phi >0$ throughout the scalar field evolution. In our proposed hilltop QI models (see Fig.~\ref{fig:QI_exp_minus_l_pot}), inflation ends for $\phi > m_p$  and thereafter the potential rapid declines upon increasing the value of $\phi$. In this case, it is difficult satisfy the non-adiabatic particle production condition (\ref{eq:QI_adiabaticity}).

        In the literature of QI featuring such potentials \cite{Ahmad:2017itq,Dimopoulos:2017tud}, the usual approach is to assume the  interaction Lagrangian to be of the  form
        
 \beq
{\cal L}_{\rm int} = - \frac{1}{2}g^2(\phi - \phi_c)^2\chi^2 - h \bar{\psi} \psi \chi~,
\label{eq:QI_L_int_modified}
\eeq       
where $\phi_c$ is a point of enhanced symmetry (thus requiring the action to be invariant under the  shift symmetry transformation $\phi \rightarrow \phi - \phi_c$). For this Lagrangian, the condition for non-adiabatic particle production becomes  \cite{Ahmad:2017itq,Dimopoulos:2017tud}

\beq
\phi_c - \sqrt{\f{|\dot{\phi}|}{g}} \,  <  \, \phi <  \, \phi_c + \sqrt{\f{|\dot{\phi}|}{g}}~,
\label{eq:QI_adiabaticity_modified}
\eeq
which can be satisfied by suitably choosing the point of enhanced symmetry to be slightly greater than the field value at the end of inflation, \textit{i.e}  $\phi_c \geq \phi_e$.  However if the Lagrangian does not possess such an enhanced symmetry, then it becomes difficult to achieve instant preheating in this scenario. Hence we propose the following novel interaction Lagrangian to alleviate this problem.

\subsection{Proposal for a new type of coupling}
\label{sec:QI_new_coupling}
As discussed above, the interaction Lagrangian
${\cal L}_{\rm int} = -\frac{1}{2}g^2\phi^2\chi^2$ has problems in dealing with
hilltop-type potentials which peak at $\phi=0$ and begin to steepen at large values of
$\phi$. Hence we propose considering a different coupling, namely
\beq
{\cal L}_{\rm int} = -\frac{1}{2}g^2\left (\frac{\rho_\phi}{\mpl^2}\right )\chi^2
\eeq
which corresponds to the effective mass 
$m^2 \equiv m^2_\chi = g^2{\rho_\phi}/{\mpl^2}$
where $\rho_\phi$ is the density of the inflaton field. So in a sense $\chi$ is 
coupled to the Hamiltonian density of $\phi$.

One can show that in this case
\beq
\frac{|{\dot m}|}{m^2} = \frac{\epsilon_H}{\sqrt{3} g}
\label{eq:adiabat}
\eeq
where
\beq
\epsilon_H = -\frac{\dot H}{H^2}
\eeq
is the slow roll parameter.

Next its easy to see that during the rapid roll of $\phi$ down a very
steep potential, $w_\phi \simeq 1$ and $\epsilon_H \simeq 3$.
Since the adiabaticity condition must be violated
during this epoch 
one finds
\beq
\frac{|{\dot m}|}{m^2} \geq 1 ~\Rightarrow~ g \leq \sqrt{3}
\label{eq:stiff}
\eeq
to be the precise requirement for preheating. However Eq.~(\ref{eq:adiabat}) also
implies another constraint, since we know that the adiabaticity condition
must be satisfied {\em during inflation} to avoid excessive particle production
during this period. This second requirement leads to
\beq
\frac{|{\dot m}|}{m^2}\Bigg\vert_{\rm Inflation} = \frac{\epsilon^{\rm inf}_H}{\sqrt{3} g} \ll 1
\label{eq:adiabat1}
\eeq
Combining Eqs.~(\ref{eq:stiff}) and (\ref{eq:adiabat1}) results in
\beq
\epsilon^{\rm inf}_H \ll \sqrt{3} g \ll \epsilon^{\rm stiff}_H
\label{eq:adiabat2}
\eeq
equivalently
\beq
\frac{r}{28} \ll g \ll \sqrt{3}
\label{eq:adiabat3}
\eeq
where we have used $\epsilon^{\rm inf}_H = r/16$, 
with $r \leq 0.036$ suggested by the latest CMB observations. Indeed substituting the upper limit $r = 0.036$
into Eq.~(\ref{eq:adiabat3}) one gets
\beq
1.3\times 10^{-3} \ll g \ll 1
\eeq
which is easy to satisfy ! Of course a lower value of $r$, as predicted in some models,
will increase the allowed range in $g$ considerably, which further adds to the appeal of this
model.

\section{Equivalence between exponential QI models}
\label{sec:app_eqv}

 In section \ref{sec:QI_dual_exp} we introduced two different  exponential based QI potentials which are discussed in sections \ref{sec:QI_dual_exp_KKLT} and \ref{sec:QI_dual_exp_lambda}. Here we demonstrate that the two models are actually equivalent and one can be derived from the other by a parameter redefinition $\frac{M}{N}\longrightarrow \lambda$.
 
\begin{enumerate}
\item  {\large Exponential plateau QI}

The plateau QI potential (\ref{eq:QI_exp_plateau_gen}) can be rewritten as 

\beq
V(\phi) = V_{\rm inf}\exp\left\lbrack-\frac{M^{2n}}{N^{2n} + \phi^{2n}}\right\rbrack~,
\label{eq:app_eqv_1}
\eeq
with $V_{\rm DE} = V_{\rm inf} \, e^{-\l(M/N\r)^{2n}} $. Expression (\ref{eq:app_eqv_1}) can also be written as 
$$V(\phi) = V_{\rm DE} \, \exp\l[M^{2n}\l(\f{1}{N^{2n}}-\f{1}{\phi^{2n}+N^{2n}}\r)\r]\, ,$$
which becomes 
$$V(\phi) = V_{\rm DE} \, \exp\l[\l(\f{M}{N}\r)^{2n} \l(\f{\phi^{2n}}{\phi^{2n}+N^{2n}}\r)\r] \, .$$

By replacing $\f{M}{N} \longrightarrow \lambda\,$, we obtain

\beq
V(\phi) = V_{\rm DE} \, \exp\l[\f{\l(\lambda\phi\r)^{2n}}{\phi^{2n}+N^{2n}}\r] \,,
\label{eq:app_eqv_2}
\eeq
which is equivalent to the plateau potential (\ref{eq:QI_exp_l_plateau_gen}).

\item  {\large Exponential hilltop QI}

The hilltop QI potential (\ref{eq:QI_exp_hilltop1}) can be written as 

\beq
V(\phi) = V_{\rm DE}\exp\left\lbrack\frac{M^{2n}}{N^{2n} + \phi^{2n}}\right\rbrack~,
\label{eq:app_eqv_3}
\eeq
with $V_{\rm inf} = V_{\rm DE} \, e^{\l(M/N\r)^{2n}} $. Expression (\ref{eq:app_eqv_3}) can be written as

$$V(\phi) = V_{\rm inf} \, \exp\l[M^{2n}\l(\f{1}{\phi^{2n}+N^{2n}}-\f{1}{N^{2n}}\r)\r] \, ,$$
which becomes
$$V(\phi) = V_{\rm inf} \, \exp{\l[-\l(\f{M}{N}\r)^{2n} \l(\f{\phi^{2n}}{\phi^{2n}+N^{2n}}\r)\r]} \, .$$

By replacing $\f{M}{N} \longrightarrow \lambda\,$, we obtain

\beq
V(\phi) = V_{\rm inf} \, \exp\l[\f{-\l(\lambda\phi\r)^{2n}}{\phi^{2n}+N^{2n}}\r] \,,
\label{eq:app_eqv_4}
\eeq
which is equivalent to the hilltop potential (\ref{eq:QI_exp_l_hilltop_gen}).
\end{enumerate}

\section{Additional potentials for Quintessential Inflation}
\label{sec:appendix}

As noted in sections \ref{sec:QI_hilltop_KKLT} \& \ref{sec:QI_dual_exp_lambda} hilltop potentials with flat wings can successfully
unify inflation with dark energy.
In this section we provide more examples of hilltop potentials and show
how they too might give helpful insights into QI model building.

\begin{itemize}

\item
A hilltop model of QI can be constructed by substituting the plateau function
\beq
f(\phi) = \left\lbrack \lambda \tanh{\left (\frac{\phi}{M} \right )} \right\rbrack^{2n}
\label{eq:tanh1}
\eeq
into 
$V(\phi) = V_0\,e^{-f(\phi)}$, while the inverse potential $V(\phi) = V_0\,e^{f(\phi)}$ describes a plateau model of
QI.

\item
Similarly
\beq
V(\phi) = V_0\,sech{\left\lbrack f(\phi)\right\rbrack}
\eeq
where $f(\phi)$ is the KKLT function \cite{Kachru:2003aw,Kachru:2003sx}
\beq
f(\phi) = \frac{({\lambda\phi})^{2n}}{M^{2n} + \phi ^{2n}}~, ~~n \geq 1~.
\eeq
describes a hilltop with flat wings.

Clearly many of the properties of hilltop potentials discussed in the main body of the paper
will be shared by the hilltop potentials discussed in this appendix.

\item
Additional hilltop potentials for QI
can be constructed by means of the ansatz

\beq
V(\phi) = \frac{V_0}{1+f(\phi)}
\label{eq:new_model}
\eeq
where $f$ is a plateau-like function, examples of which were provided in
the main body of the paper.
Such potentials generically give rise to Hilltops with flat wings.

For instance if
\beq
f(\phi) = \frac{({\lambda\phi})^{2n}}{M^{2n} + \phi ^{2n}}~, ~~n \geq 1~,
\label{eq:kklt1}
\eeq
then small values of $(\lambda\phi) \ll M$ result in the hilltop potential
\beq
V(\phi) \simeq V_0\left\lbrack 1 - \left (\frac{\lambda\phi}{M}\right)^{2n}\right\rbrack ~. 
\eeq
On the other hand, for large values of $\phi \gg M$, the potential mimicks a cosmological
constant with\footnote{Note that a small change in the function $f(\phi)$, namely
\beq
f(\phi) \to f(\phi)\times \log{\left\lbrack 1 + \left (\frac{\lambda\phi}{M}\right )^{2n}
\right\rbrack}
\eeq
leads to a late-time decline in the value of $V(\phi)$ in Eq.~(\ref{eq:new_model})
and results in a time dependent
EOS of dark energy.}

\beq
V(\phi \gg M) \simeq V_0\times\lambda^{-2n}~.
\eeq
Consequently the full potential (\ref{eq:new_model}) describes a hilltop with flat wings
which asymptotically approaches the $\Lambda$CDM model with
\beq
\rho_{\rm DE} = \frac{\Lambda}{8\pi G} = V_0\times\lambda^{-2n}~,
\eeq
at late times.
Since $V_0 \sim 10^{-10}m_p^4$ is suggested by CMB observations, and 
$\rho_{\rm DE} \sim 10^{-120}\,m_p^4$ is suggested by measurements of SNIa and BAO, one
finds 
\beq
\lambda^{2n} \sim 10^{110}
\eeq
implying that $\lambda$ is very large for $n \sim O(1)$.

It therefore follows that models of QI based on Eq.~(\ref{eq:new_model}) can only be constructed for
 large parameter values, in contrast to the models discussed earlier 
which permit successful unification between inflation and DE for moderate
values $\lambda \sim O(1)$. 

Note that for intermediate values $\lambda\phi \gg M, \, \phi \ll M$, the potential
in Eq.~(\ref{eq:new_model})
drops off as an inverse power-law
$V(\phi) \propto \phi^{-2n}$, more precisely
\beq 
V(\phi) \simeq \frac{V_0}{\lambda^{2n}}\left (\frac{M}{\phi}\right )^{2n}~.
\eeq
For large values of $n$ such a decline will be steep enough to
reheat the universe via the instant preheating mechanism discussed earlier.
Thus large values of $n$ give rise to successful preheating and more moderate values of
the parameter $\lambda$.

\end{itemize}

\printbibliography

\end{document}